\documentclass[journal npg]{copernicus}

\usepackage{amsmath,amssymb} %, amsthm} 
\usepackage[normalem]{ulem}
\usepackage[dvipsnames]{xcolor}
\usepackage{adjustbox}
\usepackage{array}
\newcommand\figurespath{figures/}
\usepackage{graphicx}% Include figure files
\usepackage[svgpath = {\figurespath}]{svg}
\graphicspath{{\figurespath}}

\usepackage{cleveref}
\usepackage{booktabs} % To thicken table lines

\begin{document}

\title{Global teleconnectivity structures of the El Ni{\~{n}}o--Southern Oscillation and large volcanic eruptions -- An evolving network perspective}

\Author[1,2]{Tim}{Kittel}
\Author[1,2]{Catrin}{Ciemer}
\Author[3]{Nastaran}{Lotfi}
\Author[3]{Thomas}{Peron}
\Author[3]{Francisco}{Rodrigues}
\Author[1,2]{J\"urgen}{Kurths}
\Author[1]{Reik V.}{Donner}

\affil[1]{Potsdam Institute for Climate Impact Research, Telegrafenberg A31, 14473 Potsdam, Germany}
\affil[2]{Department of Physics, Humboldt University Berlin, Newtonstra{\ss}e 15, 12489 Berlin, Germany}
\affil[3]{Institute of Mathematics and Computer Science, University of S\~ao Paulo, %13566-590 S\~ao Carlos, Brazil}%Department of Applied Mathematics and Statistics, 
Avenida Trabalhador Sao-carlense, 400-Centro, 13566-590 Sao Carlos, Brazil}
%\affil[4]{Department of Control Theory, Nizhny Novgorod State University, Gagarin Avenue 23, 606950 Nizhny Novgorod, Russia}
%\affil[5]{Institute for Complex Systems and Mathematical Biology, University of Aberdeen, Aberdeen AB243UE, United Kingdom}

\runningauthor{T. Kittel et al.} % if too long for running head
\runningtitle{Global teleconnectivity structures -- An evolving network perspective}

\maketitle

\begin{abstract}
\begin{nolinenumbers}
Recent work has provided ample evidence that global climate dynamics at time-scales between multiple weeks and several years can be severely affected by the episodic occurrence of both, internal (climatic) and external (non-climatic) perturbations. Here, we aim to improve our understanding on how regional to local disruptions of the ``normal'' state of the global surface air temperature field affect the corresponding global teleconnectivity structure. Specifically, we present an approach to quantify teleconnectivity based on different characteristics of functional climate network analysis. Subsequently, we apply this framework to study the impacts of different phases of the El Ni{\~{n}}o--Southern Oscillation (ENSO) as well as the three largest volcanic eruptions since the mid 20th century on the dominating spatio-temporal co-variability patterns of daily surface air temperatures. Our results confirm the existence of global effects of ENSO which result in episodic breakdowns of the hierarchical organization of the global temperature field. This is associated with the emergence of strong teleconnections. At more regional scales, similar effects are found after major volcanic eruptions. Taken together, the resulting time-dependent patterns of network connectivity allow a tracing of the spatial extents of the dominating effects of both types of climate disruptions. We discuss possible links between these observations and general aspects of atmospheric circulation.

%[Date: \today]

%\keywords{Complex networks \and Teleconnections \and El Ni{\~{n}}o--Southern Oscillation \and Volcanic Eruptions}

\end{nolinenumbers}
\end{abstract}

\introduction
\label{intro}

\begin{nolinenumbers}
The empirical analysis of climate data is fundamental to understand the evolution and develop more accurate methods for forecasting of climate phenomena like El Ni\~{n}o.
Typically, such data sets comprise time series representing temperature, precipitation or other climate variables observed at distinct locations distributed around the globe. Their common properties include long-range spatial and often also temporal correlations \citep{Fraedrich2003}, interactions at and among multiple scales \citep{Palus2014} and nonlinearity \citep{Dijkstra2013}. With the Earth's surface being subdivided into regions for which individual ``grid points'' and associated localized records of climate variability are considered representative, the evolution of the climate system can be approximately described by a high-dimensional multivariate time series composed of a multitude of interdependent signals. 

While the analysis of such big climate data sets has been traditionally attempted by means of classical statistical approaches like empirical orthogonal function or maximum covariance analysis \citep{vonStorch2003}, it has recently been realized that these methods exhibit fundamental intrinsic limitations, including their linearity and associated condition of pairwise orthogonal patterns \citep{Gamez2004}. As a consequence, the traditional view that the corresponding decompositions of global spatio-temporal co-variability patterns actually provide dynamical (or at least statistical) modes that unambiguously coincide with specific key climatic processes has been abandoned \citep{Monahan2009}. Taken together, there is growing evidence that the application of traditional linear methods of signal processing and the climatic interpretation of their results are severely affected by the dynamical complexity of the involved processes.

During the last years, complex network representations of climate variability have been developed \citep{tsonis2006networks,donges2009backbone,donges2009complex,tsonis2011community,tsonis2012,steinhaeuser2012multivariate,Peron2014,Ciemer2017} and demonstrated to provide a suitable approach for relieving some of the aforementioned concerns \citep{donges2015}. In this nonlinear statistical framework, referred to as \emph{functional climate network analysis}, the individual grid points or cells are considered as nodes of a spatially embedded graph. Connections among these nodes are established according to similarities between the individual (local) climate time series \citep{tsonis2006networks,donges2009complex,Donner2017}. By construction, the network structures thus obtained highlight essential statistical interrelationships among spatio-temporal climate data \citep{donges2009complex}. 

The application of functional climate networks has already provided several important insights. For instance, centrality measures, such as betweenness centrality, have been found to serve as tracers of global circulation patterns in the atmosphere and oceans \citep{donges2009backbone}. Moreover, climate networks have been used to identify dipole patterns which represent pressure anomalies of opposite polarity appearing in two different regions simultaneously \citep{kawale2011discovering}. The study of the coupling structure between interdependent climate variables \citep{donges2011investigating}, the temporal evolution and teleconnections of the North Atlantic Oscillation \citep{guez2012climate,guez2013global}, the distinction of different types of El Ni\~{n}o phases \citep{radebach2013disentangling,Wiedermann2016a} and the prediction of the latter \citep{ludescher2013improved,ludescher2014very} have also been subjects of corresponding recent investigations. Many of the aforementioned methodological achievements have been integrated in open source software packages \citep{donges2015chaos} contributing to the increasing use of functional network analysis in climatological studies \citep{Donner2017}.

One rather fundamental property of large networks is their (possibly hierarchical) organization in terms of communities -- an aspect that has also been addressed recently in the context of key patterns in climate data \citep{tsonis2011community}. Here, a community is a subset of densely connected nodes which exhibit only few interactions with the rest of the network \citep{newman2006modularity,fortunato2010community}. In a climate network context, communities would ideally have some climatological interpretation. Specifically, \citet{tsonis2011community} argued that each community in a climate network should be considered as a subsystem which operates relatively independent of the other communities. Besides corresponding connectivity structures in individual climate variables, community detection algorithms \citep{fortunato2010community} can also be used to detect multi-variable clusters \citep{steinhaeuser2010exploration}. 

In this paper, we analyze global surface air temperature data in terms of functional climate networks and demonstrate the close relationship between El Ni\~{n}o and La Ni\~{n}o episodes as well as strong volcanic eruptions on the one hand, and temporal changes in the modular organization of the resulting networks on the other hand. For this purpose, we study the teleconnectivity structure in the climate system in terms of spatial fields of two network properties that represent the number of strong statistical connections, as well as the average spatial distance between the connected grid points. In addition, the associated temporal variations are traced by some scalar-valued global network characteristics. 
%The earth surface is divided into Voroni cells, in which, each cell is considered as one node, and two nodes are connected if the correlation between their metric is higher than a threshold. Using the Walktrap algorithm, which is based on the fact that random walks tend to become trapped in dense part of the network corresponding to communities \cite{pons2006computing}, modularity of this climate network is calculated. The modularity measure enables a better classification for the El Ni\~{n}o and La Ni\~{n}o. With this analysis, we verify that the volcano eruption of Mount Pinatubo in 1991 influenced the evolution of El Ni\~{n}o and the network modular organization.

The remainder of this paper is organized as follows: Sect.~\ref{sec:background} provides brief information on the climatological background of ENSO and volcanic eruptions as the two types of major climatic disruptions studied in this work. The data and methodology used in this work are described in detail in Sect.~\ref{sec:methods}. Finally, our results are presented and discussed in Sect.~\ref{sec:results}, followed by concluding remarks.

~

\section{Climatological background} \label{sec:background}

\subsection{El Ni{\~{n}}o--Southern Oscillation} \label{sec:background-ENSO}

\begin{figure}
	\centering
	\includegraphics[width = \columnwidth]{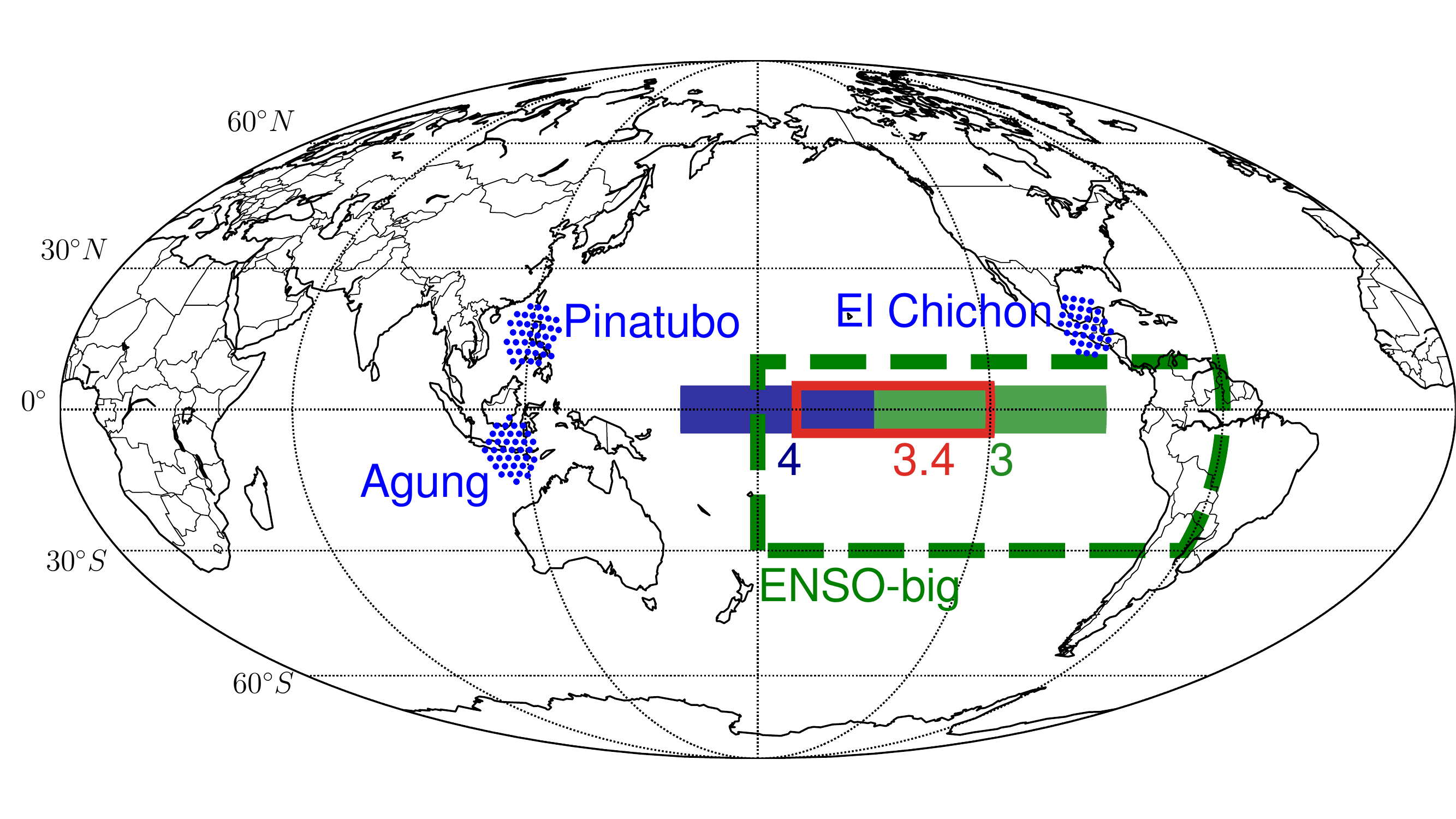}
	\caption{Main regions of interest used within this paper. Sets of blue dots labelled with ``El Chichon'', ``Agung'' and ``Pinatubo'' indicate grid points within a 5$^\circ$ radius around the corresponding volcanoes. The numbers 3, 3.4 and 4 identify the corresponding Nino regions (cf.\ Tab.~\ref{tab:enso-regions}) commonly used for defining characteristic indices of ENSO variability. The region ``ENSO-big'' will be removed from the complete global data set when analyzing the spatial imprints of volcanic eruptions to ensure that ENSO-related effects are excluded.
	}
	\label{fig:map-regions-of-interest}
\end{figure}

\begin{table}
	\begin{tabular}{l|l|l} 
		\toprule
		region & latitudes & longitudes \\ \hline
		Nino1+2 & 10\degree S - 0\degree N & 90\degree W - 80\degree W \\
		Nino3 & 5\degree S - 5\degree N & 150\degree W - 90\degree W \\
		Nino4 &  5\degree S - 5\degree N  & 160\degree E - 150\degree W \\
		Nino3.4 &  5\degree S - 5\degree N  & 170\degree W - 120\degree W \\
		ENSO-big &  30\degree S - 10\degree N  & 180\degree W - 60\degree W \\ \bottomrule
	\end{tabular}
	\caption{Overview on different regions commonly used for defining characteristic temperature-based indices associated with ENSO variability. In addition, we include the definition of the ``ENSO-big'' region studied in this work, which corresponds to the region that is discarded in our analyses of the impacts of strong volcanic eruptions on global temperature teleconnectivity.}
	\label{tab:enso-regions}
\end{table}

Among the dominant teleconnectivity patterns in the global climate system, the El Ni\~{n}o--Southern Oscillation (ENSO) is the probably most remarkable phenomenon in terms of both, the magnitude of associated variations in sea-surface temperature (SST) and sea-level pressure, as well as its specific impacts on different aspects of regional climate variability worldwide \citep{trenberth_definition_1997}. During the positive phase (El Ni\~{n}o) of this complex oscillation of the coupled atmosphere--ocean system in the tropical Pacific, the eastern tropical Pacific exhibits some anomalous warming with respect to ``normal'' mean conditions, while the negative phase (La Ni\~{n}a) is characterized by a corresponding cooling. In comparison with the normal mean climatology, this surface temperature anomaly results in marked shifts of key atmospheric pressure systems, modifying the large-scale circulation and, thus, leading to prominent shifts of, e.g., precipitation patterns. It has been shown that effects of both ENSO phases can be observed in remote regions including North and South America, Africa, the Indian subcontinent, and even Antarctica \citep{ropelewski_global_1987,dai_global_2000,neelin_tropical_2003,turner2004nino,clarke2008introduction,sarachik2010nino}. 

The long-term variability of ENSO is characterized by some irregular oscillations with a period of 2 to 7 years and remarkable variations in the associated characteristic frequencies and amplitudes of both, temperature and pressure anomalies. Following its prominent spatial structures in tropical SST and sea level pressure, ENSO is commonly traced by indices that take up the variability of the aforementioned observables in some key region of the tropical Pacific ocean. Notably, a set of indices has been defined in terms of average SST anomalies taken over distinct regions in the eastern and central tropical Pacific, referred to as Nino1+2, Nino3, Nino4 and Nino3.4, respectively \citep{trenberth2001indices} (see Fig.~\ref{fig:map-regions-of-interest} and Tab.~\ref{tab:enso-regions}). In this work, we will utilize the so-called Ocean Ni\~{n}o Index (ONI) for differentiating between different phases of ENSO. It is defined as the running three-month mean SST anomaly for the Ni\~{n}o 3.4 region ($5\degree$N--$5\degree$S, $120\degree$--$170\degree$W) in comparison with centered 30-year base periods that are updated every 5 years \citep{NCEP-ONI}. When the ONI exceeds $0.5\degree$C for at least five consecutive months, the corresponding situation is classified as an El Ni\~{n}o, and the magnitude of the ONI is taken as an indicator of the strength of the corresponding event. In turn, if the ONI drops below $-0.5\degree$C for at least five consecutive months, this indicates a La Ni\~{n}a episode.

In the last years, it has been recognized that the commonly observed spatial patterns associated with El Ni\~{n}o (as well as La Ni\~{n}a) related SST anomalies are far from being homogeneous across the set of observed events. Consequently, it has been suggested to further distinguish both phases into two respective flavours \citep[see][and references therein]{Wiedermann2016a}. The first type is the classic or East Pacific (EP) El Ni\~{n}o \citep{rasmusson1982variations,harrison1998nino}, which is localized in the eastern tropical Pacific and characterized by strong positive SST anomalies close to the western coast of South America. Opposed to this, the El Ni\~{n}o Modoki or Central Pacific (CP) El Ni\~{n}o exhibits marked SST anomalies in the central tropical Pacific close to the dateline. Notably, both spatial structures (EP and CP) can be observed in the context of La Niña, too. Noticing that there have been contradictory classifications in the literature for some past ENSO events, \citet{Wiedermann2016a} recently presented a new indicator for the ENSO flavor based on functional climate networks. In the remainder of this paper, we will follow their classification, which is summarized in Tab.~\ref{tab:summary-ep-cp-events}.

\begin{table}
	\centering
	\begin{tabular}{p{0.2\columnwidth}|p{0.7\columnwidth}}
		\toprule
		Event & Years \\ \hline
		EP El Niño & 1957, 1965, 1972, 1976, 1982, 1997 \\ 
		CP El Niño & 1953, 1958, 1963, 1968, 1969, 1977, 1979, 1986, 1987, 1991, 1994, 2002, 2004, 2006, 2009 \\ 
		EP La Niña & 1964, 1970, 1973, 1988, 1998, 2007, 2010 \\ 
		CP La Niña & 1954, 1955, 1967, 1971, 1974, 1975, 1984, 1995, 2000, 2001, 2011 \\
		\bottomrule
	\end{tabular}
	\caption{
		Classification of past ENSO episodes into the four types Eastern Pacific (EP) El Ni\~{n}o, Central Pacific (CP) El Ni\~{n}o, EP La Ni\~{n}a, and CP La Ni\~{n}a as proposed by \citet{Wiedermann2016a}. For constructing the composite maps shown in Fig.~\ref{fig:composites}, time windows corresponding to each type of event have been selected according to their midpoints coinciding with Christmas of the year given in the table. All years not listed here have been classified as ``neutral'' years with no distinct ENSO event.
	}
	\label{tab:summary-ep-cp-events}
\end{table}

\subsection{Volcanic eruptions}

Besides distinct ENSO episodes and their known global climate impacts, another type of events that can substantially affect climate at large spatial and temporal scales are strong volcanic eruptions. Similar to El Ni\~{n}o and  La Ni\~{n}a episodes, such events can result in large-scale spatially coherent cooling trends due to modifications of the radiation balance by changes in atmospheric chemistry and the shielding effect of volcanic aerosols in the stratosphere. Subsequently, such cooling can again cause changes of precipitation and temperature patterns from synoptic (weather) time scales to relatively persistent multi-annual effects \citep{Robock2000} and even trigger long-lasting climate disruptions \citep{Miller2012}. In the past decades, several large volcanic eruptions have injected up to some millions of tons of sulfur dioxide into the atmosphere, which can get rapidly distributed around the globe once entered the stratosphere. 

In this study, we focus on the global effects of the three major volcanic eruptions during the second half of the 20th century. Within this period, the largest and most influential event, the Mount Pinatubo eruption \citep{McCormick1995}, took place between April and September 1991 in the Philippines, followed by the Mount Agung eruption in Indonesia (February 1963 to January 1964) and the El Chichon eruption (March to September 1982) in Mexico (see Fig.~\ref{fig:map-regions-of-interest}).

\section{Data and methods} \label{sec:methods}

\subsection{Data}
\label{sec:data}
In this study, we use daily mean surface air temperature (SAT) data (at sigma level $\sigma=0.995$) from the National Center for Environmental Prediction (NCEP) and National Center for Atmospheric Research (NCAR) Reanalysis I project \citep{Kalnay1996,kistler_ncepncar_2001}: The data cover the years 1948--2015 at a global grid with equi-angular spatial resolution of 2.5$\degree$ in both latitude and longitude, thus comprising $10,512$ individual temperature time series. Note that we found the inclusion of the land areas important, hence we chose SAT instead of the aforementioned SST. In order to remove leading order effects of seasonality in the temperature recordings, the long-term average temperatures for each calendar day of the year have been subtracted from the raw data independently for each grid point, resulting in so-called SAT anomalies.

Equi-angular gridded data have, by construction, a higher density of grid points at the poles than around the equator, which would result in systematic biases of statistical characteristics overemphasizing the polar regions with apparently more data if not properly accounted for. For the latter purpose, area-weighted measures have been developed and subsequently applied in recent works \citep{tsonis2006networks,heitzig2012node,Wiedermann2013}. As an alternative, we follow here the approach of \citet{radebach2013disentangling}, where the original data have been remapped onto a grid with a much higher spatial homogeneity. Specifically, we use an icosahedral grid %(Fig.~\ref{fig:icosahedral-grid}), 
as described by \citet{Heikes1994}, which finally leads to a decomposition of the Earth's surface into Voronoi cells of almost the same area. In the present case, the corresponding remapping procedure results in a set of $N=$10,242 grid points that exhibit a narrowly peaked distribution of geodesic distances between direct neighbors. 

In \citet{radebach2013disentangling}, the time series associated with each new grid point have been determined based upon the values at the respective four surrounding grid points in the original equi-angular grid. In this work, we use a slightly different approach by taking the four closest points in space instead, which in some cases may deviate from the former setting. This modification is motivated by the fact that the consideration of the spatially closest ``observational'' values may provide a better approximation of climate variability at the new grid point. Moreover, these spatial neighbours can be determined efficiently using spatial search trees. Due to the commonly rather large spatial correlation length of the SAT field (as compared to other climate variables like precipitation) and its resulting spatial smoothness, we do not expect the time series resulting from both algorithmic variants to differ markedly.

Finally, we note that when using the global data set as described above, the temporal correlations associated with the key ENSO region and the surrounding parts of the Pacific ocean are known to dominate climate variability globally. This leads to undesired outcomes when aiming to resolve the effects of individual volcanic eruptions on global temperature patterns, since they might be masked by ENSO variability, especially in cases where the corresponding effects take place simultaneously with some El Ni\~{n}o (or La Ni\~{n}a) event. In fact, strong tropical volcanic eruptions have even been suggested to serve as triggers for El Ni\~no phases \citep{Khodri2017}.

In order to account for the problem of temporal co-occurrence between the effects of volcanic eruptions and ENSO events, we are going to use the full set of data when studying the effects of ENSO on global temperature teleconnectivity, while excluding the main ENSO region and its surroundings (referred to as ``ENSO-big'' in Fig.~\ref{fig:map-regions-of-interest} and Tab.~\ref{tab:enso-regions}) when studying the impacts of specific volcanic eruptions. Note that this excluded region has been chosen rather large on purpose (as an outcome of more systematic studies with variable regions to be discarded, which are not further discussed here for brevity) such as to ensure an as complete as possible separation between the direct ENSO impacts and the effects of volcanic eruptions, especially in cases of simultaneous events. In fact, when considering the full global SAT data set in the context of the impacts of volcanic eruptions, only the signatures of the Mount Pinatubo eruption are clearly visible \citep{radebach2013disentangling}. Alternative strategies for reducing the impact of ENSO variability in order to highlight the climate effects of other phenomena like volcanic eruptions might include conditioning out the effect of ONI or other representatives of the ENSO state on the local SAT variability at each grid point prior to network analysis. We outline further investigations that make use of such approaches  as a subject of future research.

\subsection{Functional climate network analysis}
\label{sec:climate-networks} 

Functional climate networks provide a coarse-grained spatial representation of the co-variability structure among globally or regionally distributed measurements of some climate variables \citep{tsonis2006networks,tsonis2012,Donner2017}. Starting from a set of records of the variable of interest, the geographical positions associated with the individual time series are identified with the $N$ nodes of some abstract network embedded on the Earth's surface. The connectivity of this network is then formed by establishing links between pairs of these nodes that fulfill some statistical similarity criterion (see below). Thus, links in such climate networks represent strong statistical associations between climate variability at different points in space. These associations may potentially indicate certain climatic processes or mechanisms interlinking the variability at the corresponding locations. Hence, the resulting linkage structure is referred to as \emph{functional connectivity}.

Like other undirected and unweighted networks, functional climate networks are conveniently represented in terms of their adjacency matrix $\mathbf{A}=(A_{ij})$, where $A_{ij} = 1$ indicates the existence of a link between node $i$ and node $j$, while $A_{ij} = 0$ corresponds to an absence of such a link. %A link can be referenced by the tuple of it's nodes $(i,j)$.
In our specific case, the matrix $\mathbf{A}$ is time-dependent, since the spatial co-variability structure of the SAT field changes with time. In such a case, we speak of an \emph{evolving climate network} \citep{radebach2013disentangling}.

\subsubsection{Network generation}

According to our considerations presented above, we take the grid points of the icosahedral grid constructed by remapping the original NCEP/NCAR reanalysis data as nodes of an evolving SAT network (i.e., we consider a fixed node set that does not change over time).

For establishing the time-dependent link set, we consider sliding windows in time covering a set of days $\{d\}=[d_0,d_0+\Delta d]$ with a width of $\Delta d=365$ days and mutual offset of 183 days between subsequent windows. Each of these windows is labeled with the corresponding midpoint date $d_{mid} = d_0 + \Delta d/2$. Our choice of 1-year windows is motivated by the fact that seasonality may not be not completely accounted for when using shorter windows, since the consideration of SAT anomalies as defined above does not exclude seasonality in the local higher-order statistical characteristics of SAT after correcting only for the mean climatology. Moreover, due to different distributions of land and water masses on both hemispheres of the Earth (with difference persistence properties of SAT), the resulting spatial co-variability structure manifested in the climate network topology may undergo seasonal variations as well, which could affect the results of our analysis presented in this work.

From the seasonally adjusted temperature data at each grid point during a time window (corrected for the window-wise mean), $T_i(\{d\})$ (with $i$ denoting the respective grid point), we compute the matrix of pairwise Pearson's correlation coefficients
\begin{equation}
\label{eq:correlation-coefficient}
c_{ij}(d_{mid}) = \frac{\left<T_i(\{d\})T_j(\{d\})\right>_{\{d\}}}{\sqrt{\sigma_{\{d\}}(T_i(\{d\}))\cdot\sigma_{\{d\}}(T_j(\{d\}))}}\,,
\end{equation}
\noindent
where $\left<\bullet\right>_{\{d\}}$ and $\sigma_{\{d\}}(\bullet)$ denote the mean value and standard deviation of the respective variable taken over the time window $\{d\}$. From this matrix, we identify the entries (i.e., pairs of nodes) with the highest absolute values of mutual correlations $|c_{ij}|$. Specifically, in this work, we consider the 0.5\% strongest pairwise statistical similarities among all nodes per window, i.e.,
\begin{equation}
\label{eq:adjacency-matrix}
A_{ij}(d_{mid}) = \Theta(|c_{ij}|(d_{mid})-q_{|c|,0.995}(d_{mid}))-\delta_{ij}\,,
\end{equation}
\noindent
where $\Theta(\bullet)$ is the Heaviside function, $q_{|c|,0.995}(d_{mid})$ is the $99.5$-percentile of the distribution of absolute correlation values for the time window centered at $d_{mid}$, and $\delta_{ij}$ denotes Kronecker's Delta. $\mathbf{A}\left(d_{mid}\right)$ is now the mathematical representation of our evolving climate network. In the following, we omit the explicit time dependence of $\mathbf{A}$ (also in all network properties) for brevity. Note that according to our construction, $\mathbf{A}$ is symmetric.%(corresponding to an undirected network).

\subsubsection{Node degree}

The degree $k_i$ of a node $i$ is defined as the number of links connected to $i$,
\begin{equation}
k_i = \sum_{j=1}^N A_{ij}\,.
\end{equation}
\noindent
It represents how densely a node is connected within the network. In case of a functional climate network, the degree can thus be considered as a proxy for the importance (or centrality) of a certain grid point in the spatio-temporal correlation structure of the variable of interest. 

In the following, we refer to network measures like the degree, which provide a characteristic value specific to each individual node $i$, as local network characteristics. We call the full set of their values taken together with the associated spatial positions of all nodes a \emph{field}.

\subsubsection{Average link distance}

Another local network characteristic that defines a field of values upon a functional climate network is the average link distance \citep{Donner2017} of a node $i$, which is defined as
\begin{equation}
d_i = \left<d_{ij}\right>_{\left\{j|A_{ij}=1\right\}}= \frac{1}{k_i} \sum_{\left\{j|A_{ij}=1\right\}} d_{ij}= \frac{1}{k_i} \sum_{j=1}^N A_{ij}D_{ij}\,,
\end{equation}
\noindent
with $d_{ij}$ being the normalized spatial distance between two nodes $i$ and $j$. As a proper normalization, we choose here the largest possible (shortest) distance between two points on the Earth's surface, i.e., half of the circumference of the Earth, $u_{Earth}$, so that $d_{ij} = 2 D_{ij}/u_{Earth}$ where $D_{ij}$ is the geodesic distance between nodes $i$ and $j$. A low average link distance indicates that $i$ has very localized connections, while a high value points to a node with long-ranging spatial connections. This measure is closely related with the \emph{total distance} of a node with respect to the rest of the network as previously used by \citet{Tsonis2008} in the context of functional climate networks.

We emphasize that the average link distance must not be confused with the conceptually related average path length, where $d_{ij}$, as defined above, would be replaced by the minimum number $l_{ij}$ of links separating two nodes $i$ and $j$ in the network (i.e., where $i$ and $j$ do not need to be directly connected).

Taking the average of $d_i$ over all nodes $i$ of the network gives the \emph{global average link distance}
\begin{equation}
	\left<\left<d\right>\right> = \left<d_i\right>_i .
\end{equation}
\noindent

\subsubsection{Transitivity}

The network transitivity quantifies how strongly the connectivity of a network is clustered. Mathematically, it describes the degree to which the network's adjacency property is transitive, i.e., the fraction of cases in which the presence of two links between nodes $i$ and $j$ as well as $i$ and $k$ is accompanied by a third link between $j$ and $k$. Mathematically, this is expressed as \citep{boccaletti2006,radebach2013disentangling}
\begin{align}
\mathcal{T} = \frac{\sum_{i,j,k=1}^N A_{ij} A_{jk} A_{ki}}{\sum_{i,j,k=1}^N A_{ij} A_{jk}}\, .
\end{align}
\noindent
Like the global average link distance $\left<\left<d\right>\right>$ (but unlike degree and average link distance), $\mathcal{T}$ does not define a field, but returns one single single scalar value for each network.

%The measure is how much a graph (or its sub-graphs) are clustered \citep{saramaki2007generalizations,heitzig2012node}. This is done by describing the following probability. If one chooses randomly a node and two random links, what is the probability that the two adjacent nodes are connected, too, i.e. there is a triangle? \cite{Wiedermann2016a} used this in the context of functional climate networks to create an indicator for discriminating EP and CP El Niño and La Niña events. The transitivity measure is put in equation as follows

\subsubsection{Modularity}

The concept of modularity was introduced into network science by \citet{newman2004finding} to measure the degree of heterogeneity within the network structure, i.e., how well different groups of nodes can be distinguished that are densely connected within each group, but only sparsely among different groups. In the case of a climate network, modularity provides a single scalar-valued characteristic property that discriminates between a relatively homogeneous link placement (low modularity) and the existence of (commonly regionally confined) clusters of nodes (time series) that exhibit relatively coherent variability (high modularity).

The definition of modularity relies upon a partitioning of the network into meaningful subgraphs. Up to a multiplicative constant, it counts how many links are clustered within these subgraphs and compares this value with the expected number of links inside these subgraphs if the network were random, 
\begin{equation}
\mathcal{Q} = \frac{1}{2m} \sum_{ij}\left( A_{ij} -\frac{k_i k_j}{2m}\right)\Delta_{ij}\,, \label{eq:modularity}
\end{equation}
where $m$ is the total number of links and $\Delta_{ij}$ an indicator function informing whether or not two nodes $i$ and $j$ belong to the same subgraph in the considered partition. %The implicit dependence on $d_{mid}$ has been omitted here again.
%In case of of a partition in two parts only, this equation is often written as 
%\begin{equation}
%\mathcal{Q} = \frac{1}{4m} \sum_{ij}\left( A_{ij} -\frac{k_i k_j}{2m}\right)s_i s_j \label{eq:modularity-2-groups}
%\end{equation}
%where $s_i = 1$ if $i$ belongs to group 1 and $s_i = -1$ if $i$ belongs to group 2 (see \cite{newman2006modularity}). In the derivation of \Cref{eq:modularity-2-groups}, the identity $2m = \sum_i k_i = \sum_{ij} A_{ij}$ has been used.

The individual subgraphs in the partition that maximizes the modularity $\mathcal{Q}$ are called communities. The higher the modularity, the more split-up (or modular) the network. Accordingly, community detection by modularity maximization has become a common tool for cluster analysis.

While the above definition of modularity is mathematically precise, its maximization is a hard computational problem and can only be achieved by making use of suitable heuristics. Various estimation algorithms have been proposed \citep{fortunato2010community}. It should be emphasized that many of them can result in suboptimal solutions. Thus, a good choice of the algorithm is important for obtaining reliable results. In this work, we employ the \textit{WalkTrap} method introduced by \citet{pons2006computing}. By comparing the results provided by this algorithm with those of other methods (cf. \Cref{app:modularity-comparison}), we have found that the WalkTrap solution exhibits comparably high values of modularity and relatively stable values in case of strongly overlapping windows (i.e., in cases where the considered data do not change much).

%A further analysis of the algorithms we compared can be found in suppl. \ref{suppl:cmp-communities}.

\subsection{Regionalization of field measures}
\label{sec:localization-of-field-measures}

As detailed above, node degree and average link distance constitute two important local network characteristics. In some of our following investigations, it will be useful to study the associated spatial fields in full detail. However, when focusing on the specific impacts of certain climate phenomena, it can be beneficial to perform a regionalization of these measures. Specifically, for a subset of nodes $\mathcal{X} \subseteq \left\{ 1, \dots, N \right\}$ representing a certain part of the globe, a regionalized version of the degree would be given as
\begin{equation}
k_{\mathcal{X}} = \frac{1}{|\mathcal{X}|}\sum_{i \in \mathcal{X}} k_i\,, \label{eq:localization-procedure}
\end{equation}
where $|\mathcal{X}|$ denotes the number of nodes in the considered set. As a consequence, we can not only assign a degree value to an individual node, but also (as a mean degree) to a subgraph. Note that this regionalized degree differs from the concepts of cross-degree and cross-link density between subgraphs \citep{donges2011investigating}, since unlike $k_{\mathcal{X}}$, the latter exclude contributions due to links between nodes within $\mathcal{X}$ in their definition.

For the average link distance, the corresponding regionalized property $d_{\mathcal{X}}$ can be defined in full analogy. 

Below, we detail some reasonable choices for $\mathcal{X}$ to be utilized in the context of the present work, which focus on specific spatially contiguous regions of the Earth's surface that are associated with ENSO or volcanoes with strong past eruptions.

\subsubsection{El Ni{\~{n}}o--Southern Oscillation regions}
\label{sec:localization-ENSO}

As already detailed in Sect.~\ref{sec:background-ENSO}, there exist a variety of indices that measure the ``strength'' of a particular ENSO state. Among others, four Nino regions (1+2, 3, 4 and 3.4) have been defined to capture SST anomalies in different parts of the tropical Pacific.

The regionalization approach described above can be applied to these four regions by taking all nodes located within the respective spatial domains and apply Eq.~\eqref{eq:localization-procedure}. Thereby, we obtain a set of eight new scalar-valued characteristics: $k_{\text{Nino1+2}}$, $d_{\text{Nino1+2}}$, $k_{\text{Nino3}}$, $d_{\text{Nino3}}$, $k_{\text{Nino4}}$, $d_{\text{Nino4}}$, $k_{\text{Nino3.4}}$ and $d_{\text{Nino3.4}}$. In order to reduce this vast amount of information, in what follows, we will not make use of the (anyway less frequently studied) Nino1+2 region, but focus on the Nino3.4 region (which is also the basis of the nowadays most common ONI index) and its two contributors, Nino3 and Nino4.%, the spatial overlap of which defines the Nino3.4 region.

\subsubsection{Volcano regions}
\label{sec:localization-volcanoes}

The locations of the three volcanoes responsible for the largest eruptions of the recent decades are shown in Fig.~\ref{fig:map-regions-of-interest}. To obtain interpretable information on the (tele-) connectivity induced by these eruptions, we need to integrate the connectivity properties of a sufficiently large amount of meaningfully chosen grid cells. As a first attempt, we therefore take the area within a radius of $5\degree$ around each volcano as basis for the regionalization procedure of $k_i$. This leads to the three observables $k_{\textrm{Pinatubo}}$, $k_{\textrm{Agung}}$ and $k_{\textrm{Chichon}}$. For the average link distance, one could again proceed in a similar way.

However, the aforementioned choice might not be optimal, since symmetric spatial regions in the near-field do not necessarily exhibit the strongest persistent temperature effects after an eruption. Instead, the specific local meteorological conditions (especially wind fields) during the eruption period largely control the three-dimensional patterns of atmospheric aerosol concentrations and, hence, the position of the strongest mid-term cooling to be expected. Accordingly, the induced teleconnectivity can be more evident within regions that have been shifted with respect to the locations of the volcanoes. To account for this, we also calculate regionalized degrees for accordingly shifted regions (see Sect.~\ref{sec:results-volcanoes} for details), denoted as $k^\prime_{\textrm{Pinatubo}}$, $k^\prime_{\textrm{Agung}}$ and $k^\prime_{\textrm{Chichon}}$. Here, the specific regions have been selected according to an examination of the resulting degree fields of the SAT networks for time windows succeeding the individual eruptions and the corresponding wind fields, seeking for the timing and position of the strongest anomalies in the degree field that could be attributed to each eruption (see below). Note that although a volcanic eruption may start relatively abruptly, its larger-scale atmospheric effects commonly become effective only with a considerable delay of several months or more \citep{Robock2000,McCormick1995}.

\section{Results}
\label{sec:results}

%The presentation of the results with the accompanying discussion is structured in two parts. First, we analyze the impacts of the El Ni{\~{n}}o--Southern Oscillation (ENSO) using the aforementioned global and localized observables. Afterwards, we will discuss the global SAT signatures of Mount Pinatubo's, Mount Agung's and El Chichon's volcanic eruptions.

In the following, we present the results of our functional network analysis of global SAT patterns with a focus on the associated imprints of ENSO. Subsequently, we turn to analyzing and discussing the excess connectivity induced by strong volcanic eruptions.

\subsection{El Ni{\~{n}}o--Southern Oscillation}
\label{sec:results-ENSO}

\newlength{\timeserieslen}
\setlength\timeserieslen{0.9\textwidth}

% global observables

Let us start with investigating the global effects of ENSO on the spatio-temporal co-variability structure of global SAT. From a complex network perspective, this problem has already been addressed in a variety of previous studies \citep[e.g.][and various others]{tsonis2008prl,yamasaki2008,gozolchiani2008,radebach2013disentangling,Wiedermann2016a,Fan2017}, making use of different approaches for constructing network structures from global climate data. However, none of these works has considered the complementarity between topological and spatial network properties in great detail, nor utilized the concepts of modularity and global average link distance that constitute key aspects of this paper and provide important new insights as demonstrated in the following.

\subsubsection{Global network properties}

\begin{figure*}
	\centering
	\includegraphics[width=0.8\textwidth]{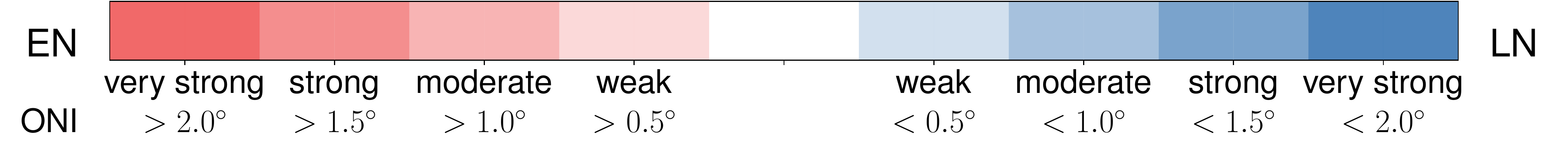} \\
	\subfloat{\includegraphics[width = \timeserieslen]{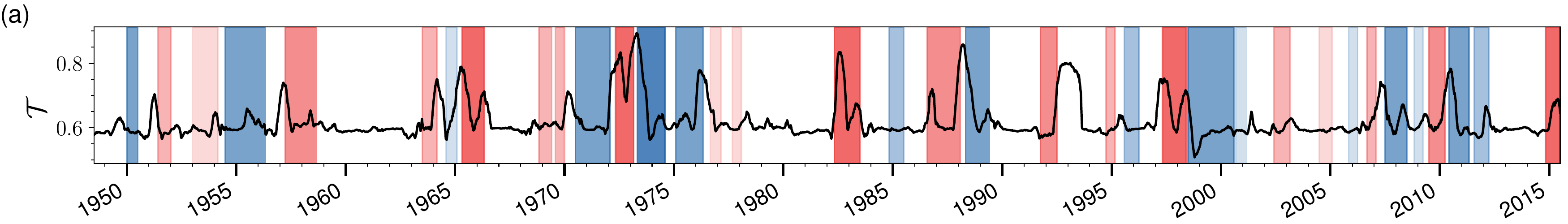}\label{fig:ts-global-transitivity}} \\
	\subfloat{\includegraphics[width = \timeserieslen]{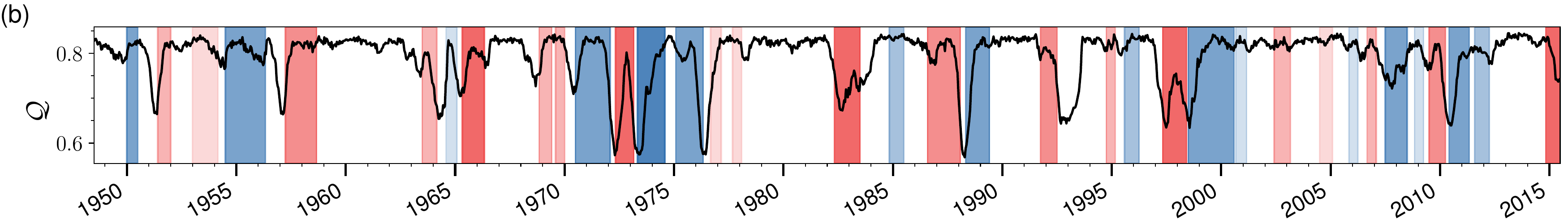}\label{fig:ts-modularity}} \\
	\subfloat{\includegraphics[width = \timeserieslen]{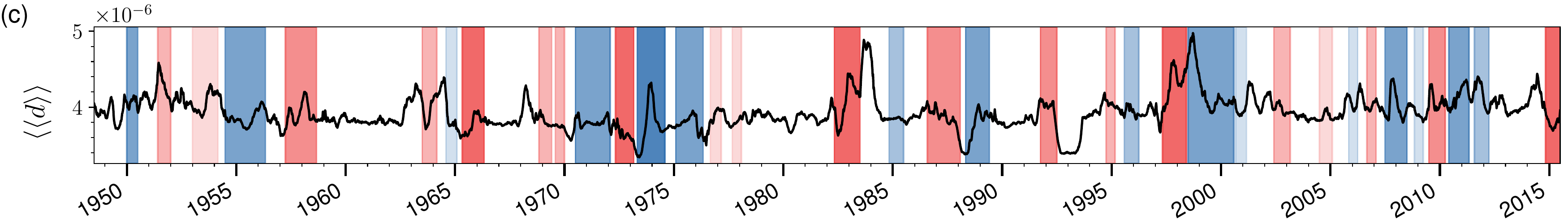}\label{fig:ts-global-link-length}}
	\caption{Time series of climate network (a) transitivity $\mathcal{T}$, (b) modularity $\mathcal{Q}$ and (c) global average link distance $\left<\left<d\right>\right>$. Background colors highlight different ENSO phases (red: El Ni\~{n}o (EN), blue: La Ni\~{n}a (LN)) according to the Ocean Ni\~no Index (ONI), with opacity representing to the corresponding index value. Ticks on the time axis indicate the 1st of January of a given year; all values are shown according to the midpoint dates of the respective time windows.}
	\label{fig:ts-global}
\end{figure*}

The network transitivity $\mathcal{T}$ has been shown by \cite{Wiedermann2016a} to systematically discriminate between the EP and CP flavours of both, El Ni\~no and La Ni\~na. While this reference used an area-weighted version of $\mathcal{T}$ and included information on the total pairwise correlation strength instead of just binary adjacency information, we follow here the approach of \citet{radebach2013disentangling} in using remapping onto an icosahedral grid. Figure~\ref{fig:ts-global-transitivity} shows the corresponding results obtained using our slightly modified data set, which are qualitatively almost indistinguishable from those of the two aforementioned studies as expected\footnote{Note, however, that Fig.~\ref{fig:ts-global} shows the results for different network measures in dependence on the window midpoint, while \citet{Wiedermann2016a} used the endpoint, leading to a 6-month shift between the respective plots.}.

In order to further quantify the strength of teleconnectivity in the global SAT field, the network modularity $\mathcal{Q}$ provides a prospective candidate measure that has not yet been exploited for this purpose in previous studies. Recall that a high modularity indicates a fragmented network, whereas low values would point to a relatively homogeneous connectivity structure of the network as a whole. Hence, a marked decrease in modularity could indicate an increase in the degree of organization of the global SAT network, i.e., a tendency towards more balanced co-variability in global temperatures.

Figure~\ref{fig:ts-modularity} shows that most time intervals that are characterized by elevated values of network transitivity actually exhibit a marked reduction in modularity. Consistent with previous findings of \citet{radebach2013disentangling}, most of these time windows in fact coincide with either some El Ni\~no or La Ni\~na phase, indicating again the global impact of these episodes in terms of equilibrating spatial co-variability in the Earth's SAT field. This can be considered as an expected signature of emerging teleconnectivity. Note that taken alone, this process would not necessarily imply a stronger \emph{synchronization} (as studied by, e.g., \citet{Maraun2005}) between climate variability in distinct regions, which would be reflected by higher absolute correlation values. Specifically, in this work, we use a fixed link density of 0.5\% in all window-specific climate networks and thus cannot make any statements about the overall strength of correlations. However, following previous results by \citet{radebach2013disentangling}, we may actually expect that the correlation threshold $q_{\left|c\right|,0.995}$ used for establishing network connectivity in this work exhibits maxima whenever $\mathcal{T}$ shows a peak, thereby supporting the hypothesis of El Ni\~no and La Ni\~na episodes synchronizing global SAT variability by establishing teleconnections.

Regarding the latter observation, it is remarkable that previous works by other authors rather reported a reduction of connectivity associated with a breakdown of synchronization due to the large-scale climate disruption triggered by El Ni\~no events \citep{yamasaki2008}. In fact, this observation has been recently used to develop network-based forecasting strategies for El Ni\~no \citep{ludescher2013improved,ludescher2014very}. However, the apparent contradiction between the latter results and our observations can be resolved when taking the different approaches of network construction used in the respective works into account.
 
Beyond their overall large-scale similarity, the temporal variability profiles of transitivity and modularity also exhibit some important differences. In particular, the strong 1982/83 El Ni\~no is represented as a single long episode of reduced modularity values while being split into two rather distinct peaks in transitivity (see Fig.~\ref{fig:ts-global-transitivity} and \ref{fig:ts-modularity}). Given the known seasonal profile of El~Ni\~no peaking around Christmas, it is remarkable that the ONI index has remained high during a quite long period of time, indicating a single extended event even despite its temporary decay captured by $\mathcal{T}$, but not quite by $\mathcal{Q}$. This underlines that both measures actually capture different aspects of network organization that provide complementary information. 

Another notable observation is related with the abrupt shift from El Ni\~no to La Ni\~na conditions in summer 1998, leading to a very fast reorganization of the global SAT field. The latter transition is reflected by some negative anomaly of $\mathcal{T}$ in summer 1998 in comparison with the ``normal'' background values of this measure, which presents a unique feature in the time evolution of network transitivity over the last decades that is not accompanied by any corresponding anomaly in $\mathcal{Q}$.

Taken together, modularity and transitivity evolve similarly at larger time scales, but provide complementary viewpoints. High transitivity commonly coincides with the temporary appearance of densely connected structures in the functional climate network, which are typically well localized in space \citep{radebach2013disentangling}. In turn, modularity captures the global connectivity pattern rather than primarily local features. Specifically, a low modularity value actually highlights more \emph{global connections} in the climate network.

While network transitivity and modularity present two key topological network characteristics, functional climate networks are systems embedded in geographical space. Thus, the spatial placement of nodes and links (which is disregarded by topological characteristics) can play a pivotal role in network structure formation \citep{radebach2013disentangling}. In order to address this aspect, we present the temporal evolution of the global average link distance $\left<\left<d\right>\right>$ in Fig.~\ref{fig:ts-global-link-length}. Notably, this measure exhibits more irregular variability with a less clear distinction between ``background level'' and ``anomalies'' associated with different types of climate disruptions than the previously studied two topological characteristics. Yet, the general behaviour of $\left<\left<d\right>\right>$ resembles that of network transitivity in the sense that ENSO-related peaks often co-occur in both measures. This indicates that strong El Ni\~no and La Ni\~na episodes do not exclusively trigger short-range (localized) connectivity (high $\mathcal{T}$), but also global teleconnectivity (high $\left<\left<d\right>\right>$), which is in line with contemporary knowledge on the large-scale impacts of both types of ENSO phases. Notably, this result is in agreement with previous qualitative results of \citet{radebach2013disentangling} on the link distance distribution of global SAT networks.

From the results discussed above, we tentatively conclude that in order to distinguish globally influential ENSO events from episodes of minor (or more regional) relevance, a combination of modularity and average link distance can be useful, taking a holistic view in studying the differential imprints of different types of ENSO phases. We will recall this strategy when discussing the effects of volcanic eruptions on network organization at a global scale.

\subsubsection{Spatial patterns of network connectivity}
% global composites

\begin{figure*}
	\centering
	\newlength\figurewidth
	\setlength\figurewidth{0.4\textwidth}
	\newlength\figuredistance
	\setlength\figuredistance{0.08\textwidth} 
    \includegraphics[width=\figurewidth]{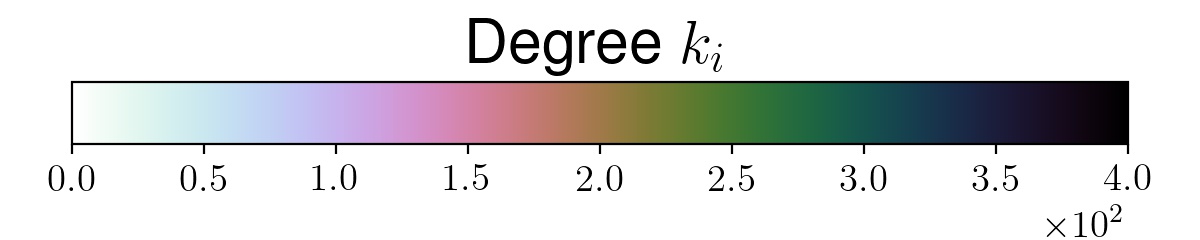} \hspace{\figuredistance} \includegraphics[width=\figurewidth]{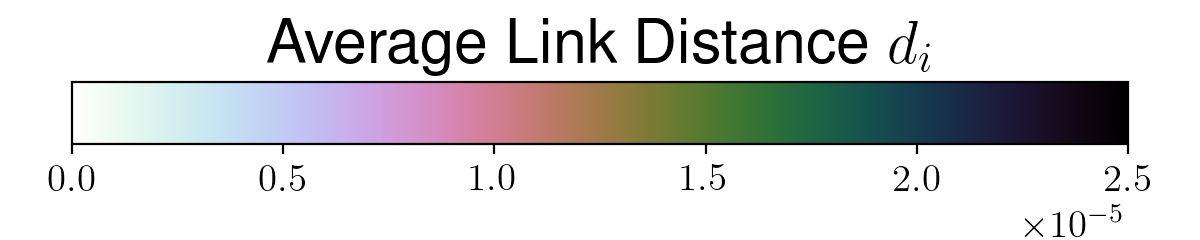} \\
	\subfloat{\label{fig:composite-marc-en-ep-degree}\includegraphics[width=\figurewidth]{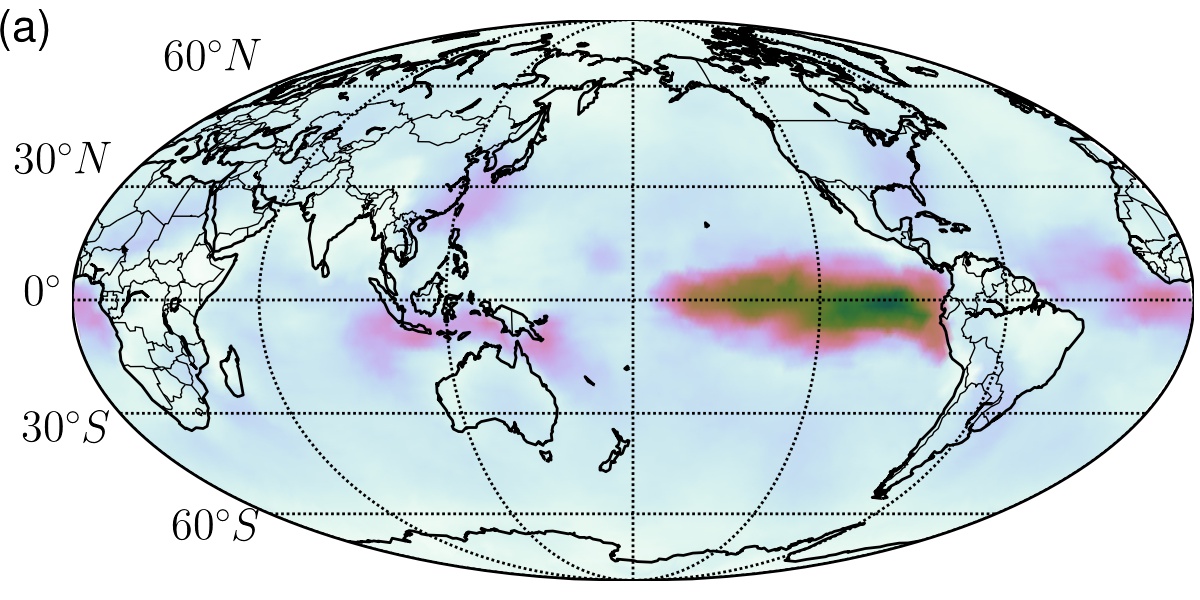}} \hspace{\figuredistance} \subfloat{\label{fig:composite-marc-en-ep-link-length}\includegraphics[width=\figurewidth]{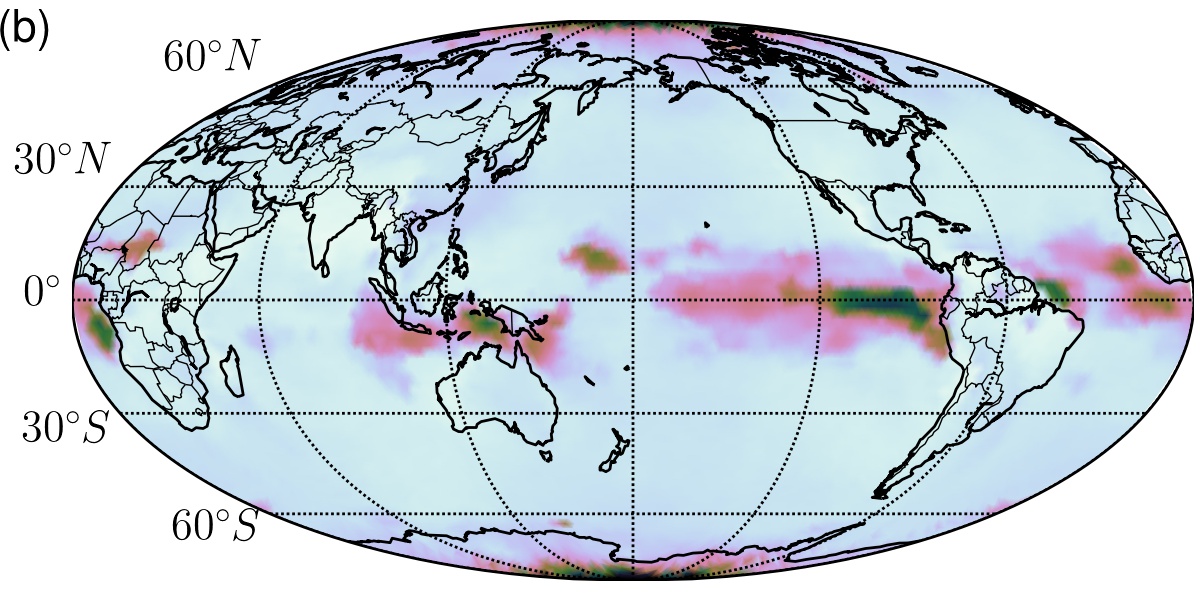}}  \\
	\subfloat{\label{fig:composite-marc-en-cp-degree}\includegraphics[width=\figurewidth]{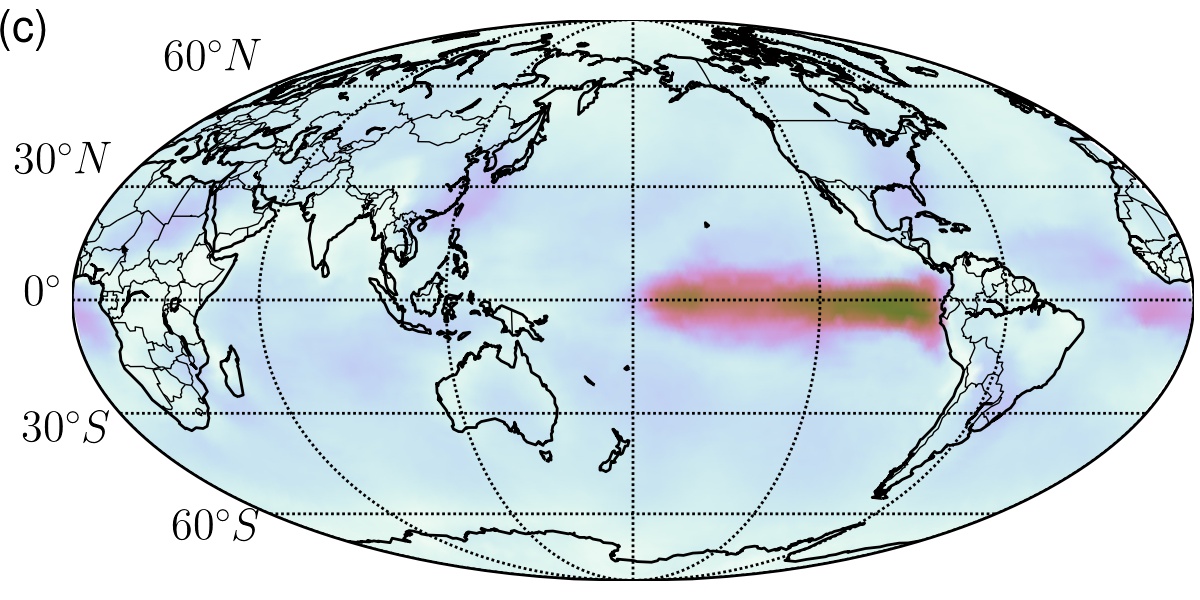}} \hspace{\figuredistance} \subfloat{\label{fig:composite-marc-en-cp-link-length}\includegraphics[width=\figurewidth]{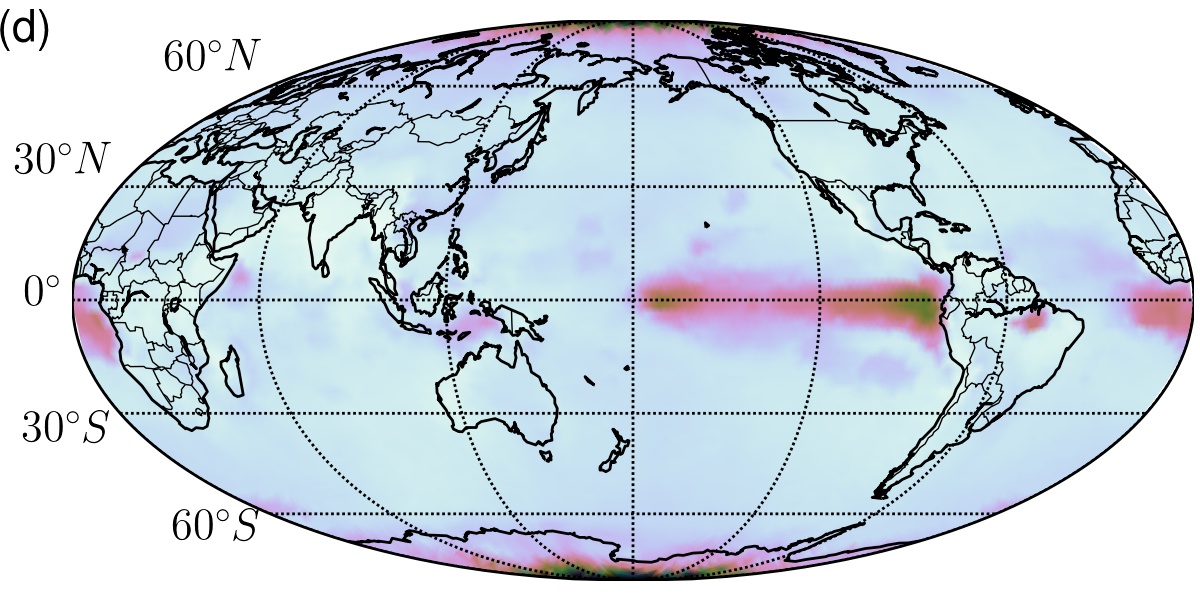}}  \\
	\subfloat{\label{fig:composite-marc-ln-ep-degree}\includegraphics[width=\figurewidth]{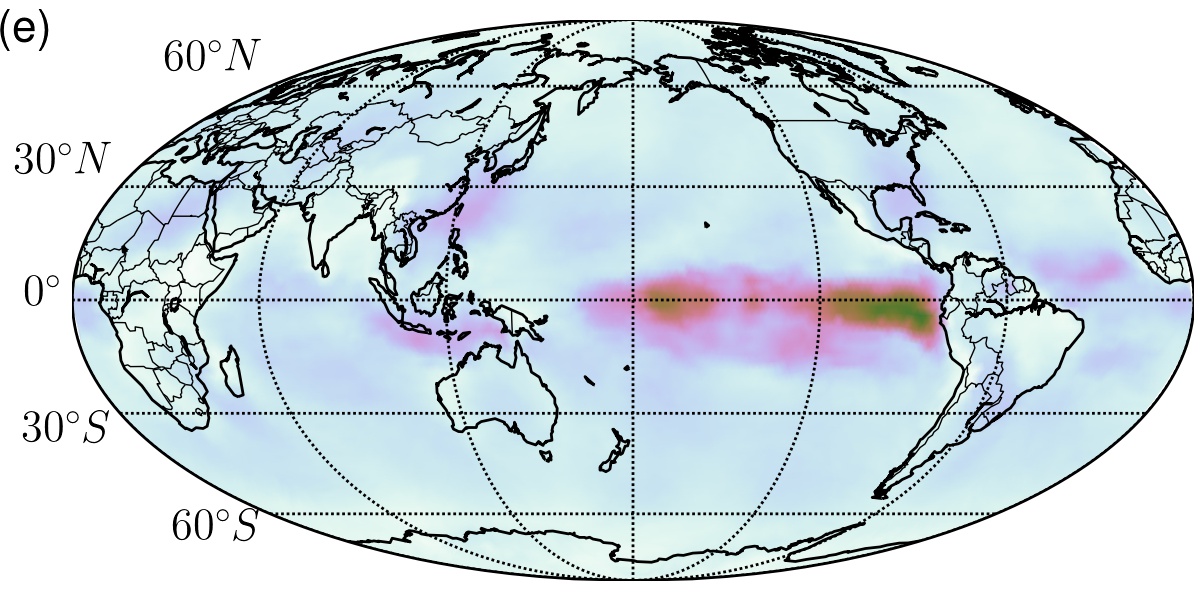}} \hspace{\figuredistance} \subfloat{\label{fig:composite-marc-ln-ep-link-length}\includegraphics[width=\figurewidth]{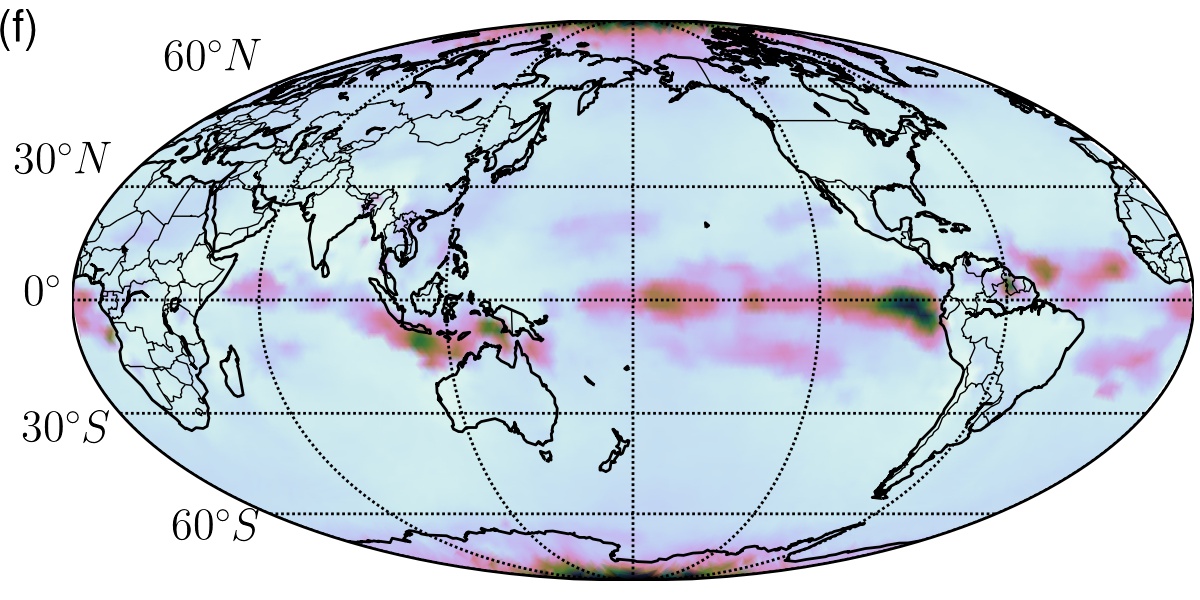}}  \\
	\subfloat{\label{fig:composite-marc-ln-cp-degree}\includegraphics[width=\figurewidth]{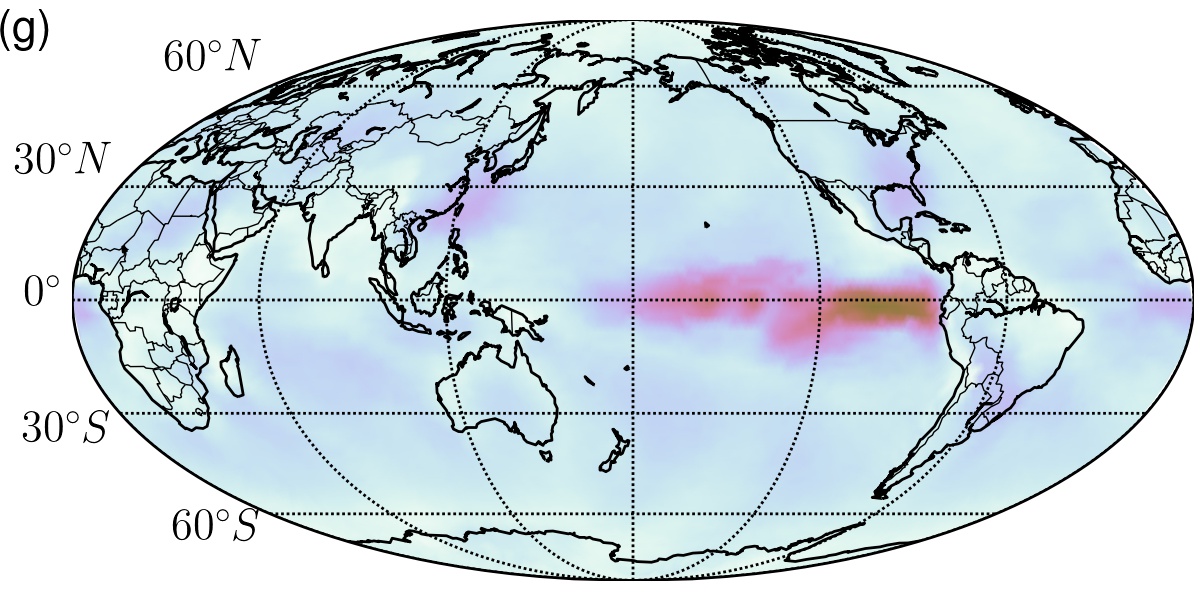}} \hspace{\figuredistance} \subfloat{\label{fig:composite-marc-ln-cp-link-length}\includegraphics[width=\figurewidth]{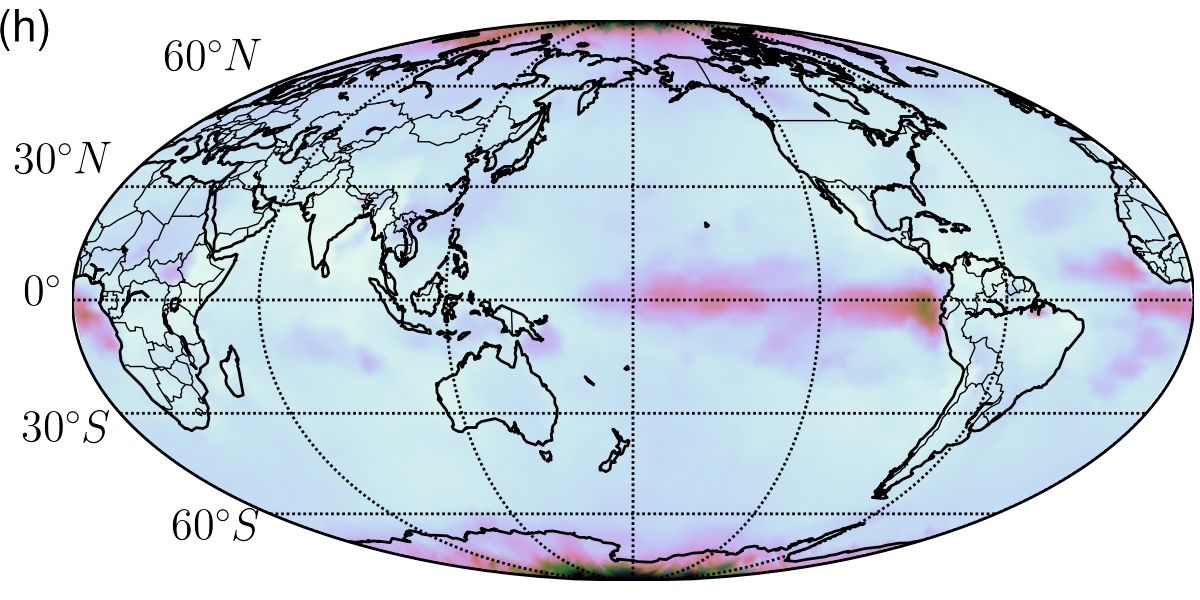}}   \\
	\subfloat{\label{fig:composite-other-degree}\includegraphics[width=\figurewidth]{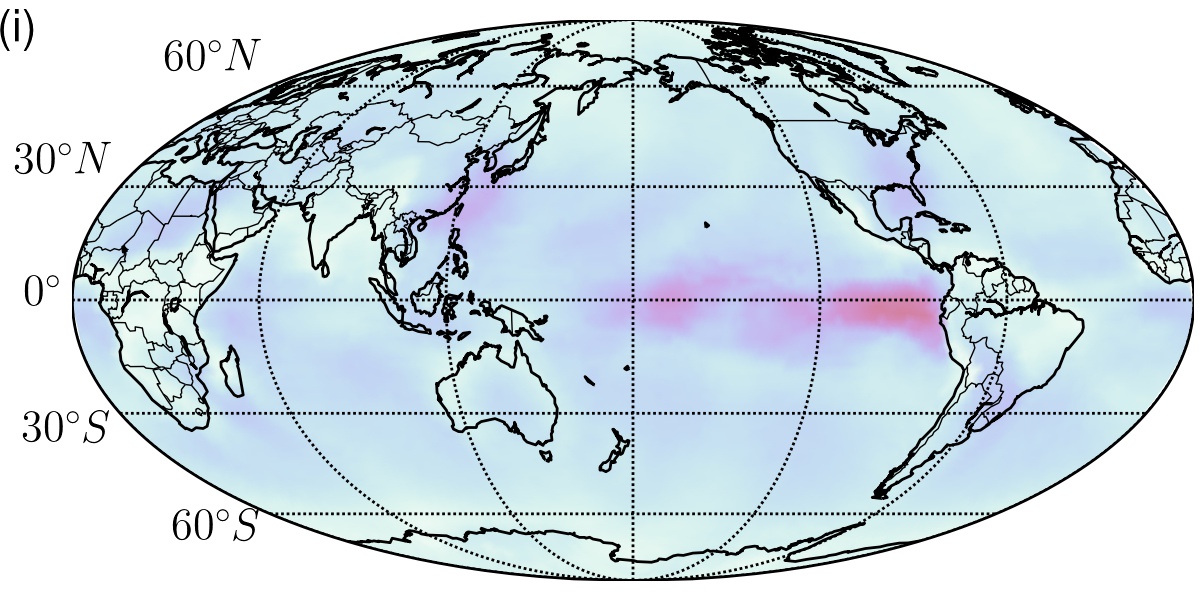}} \hspace{\figuredistance} \subfloat{\label{fig:composite-other-link-length}\includegraphics[width=\figurewidth]{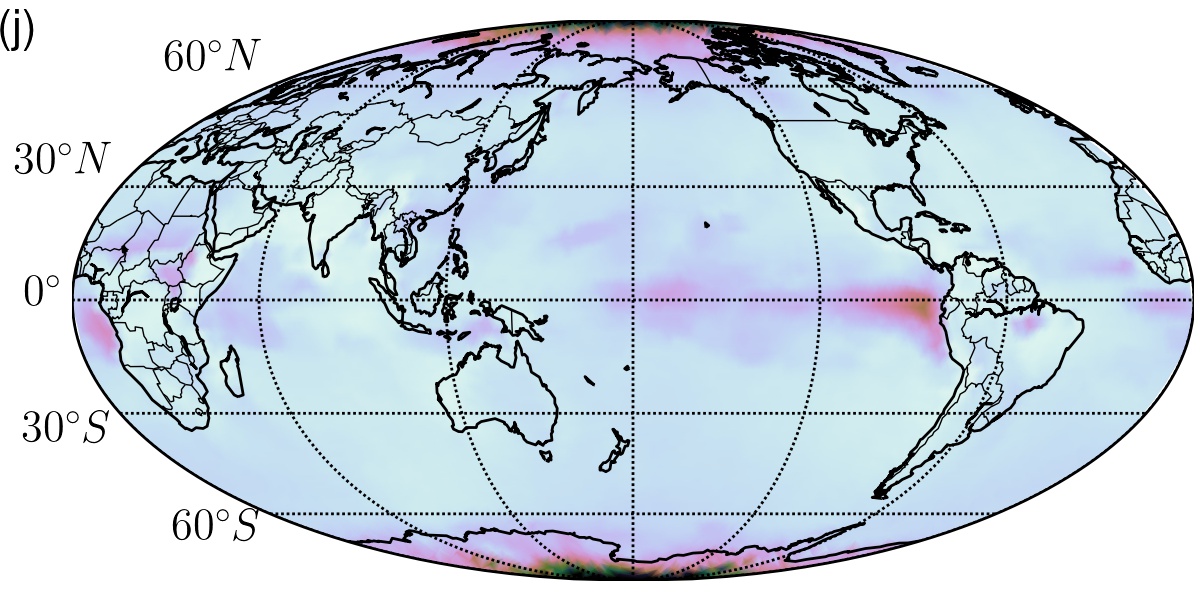}}  
	\caption{Global maps showing composites of (a,c,e,g,i) degree $k_i$ and (b,d,f,h,j) average link distance $d_i$  for different types of ENSO phases:
(a,b) EP El Niño, 
(c,d) CP El Niño, 
(e,f) EP La Niña, 
(g,h) CP La Niña 
and (i,j) all other periods. The corresponding classification of different years is summarized in Tab.~\ref{tab:summary-ep-cp-events}. 
%Generally, the plots in the left column indicate that the different ENSO phases have a strong effect within the climate network. Further, the right columns add to this interpretation that these effects include teleconnectivity. A detailed analysis is in \Cref{sec:results-ENSO}.
}
\label{fig:composites}
\end{figure*}

The above analysis of global network properties has largely confirmed some known effects of certain ENSO phases on the spatial co-variability structure of the global SAT field. Drawing upon the insight that topological and spatial network measures can provide different perspectives on the corresponding network patterns, we now turn to investigating the geographical characteristics of the generated functional climate networks. Specifically, following recent observations that climate network properties distinguish between the EP and CP flavours of both, El Ni\~no and La Ni\~na \citep{Wiedermann2016a}, we are interested in the question how the associated (tele-) connectivity structures are manifested in the respective spatial fields of degree and average link distance. For this purpose, Fig.~\ref{fig:composites} shows composite plots of the spatial patterns exhibited by both network properties during the different types of ENSO phases, thereby averaging the local network properties over all time windows that are classified as showing either of these situations (see Tab.~\ref{tab:summary-ep-cp-events}). 

%As representatives for the field for each event, we used the values for Christmas (Dec 24) of the corresponding year because this is the usual time where El Niño / La Niña have their peak intensity.

The left panels of Fig.~\ref{fig:composites} display the respective mean degree fields for the different types of ENSO periods. 
As expected, we observe a particularly strong deviation from a homogeneous pattern during EP El Ni\~nos (Fig.~\ref{fig:composite-marc-en-ep-degree}), while the degrees in the eastern-to-central tropical Pacific are only slightly larger than in the rest of the network during time windows without El Ni\~no or La Ni\~na conditions (Fig.~\ref{fig:composite-other-degree}). This general behaviour is expected from previous studies \citep{Wiedermann2016a}. Still, the observed degree patterns alone do not allow us to distinguish between a local or global phenomenon. For this purpose, the right panels of Fig.~\ref{fig:composites} show the corresponding mean average link distance fields for each type of situation. Elevated values of this measure in the typical ENSO region are present in case of all four possible types of episodes, indicating that both flavours of El Ni\~no and La Ni\~na actually generate additional connections in the tropical Pacific that span relatively large distances.

Analyzing the composite maps of the average link distance in more detail, it is important to note that beyond the ENSO region itself, additional parts of the globe exhibit elevated values. This indicates the presence of localized teleconnections that possibly link climate variability in the latter regions with ENSO. Specifically, EP El Ni\~nos (Fig.~\ref{fig:composite-marc-en-ep-link-length}) exhibit such teleconnections with Indonesia and Western Africa, which are also recovered for EP La Ni\~nas (Fig.~\ref{fig:composite-marc-ln-ep-link-length}). For CP El Ni\~nos (Fig.~\ref{fig:composite-marc-en-cp-link-length}), the $d_i$ field highlights a weak connection with Western Africa, but none with Indonesia. Similar but still weaker teleconnections can be observed for CP La Ni\~nas (Fig.~\ref{fig:composite-marc-ln-cp-link-length}). 

Among the aforementioned patterns, the apparent teleconnection with Indonesia present during EP events but not during their CP counterparts is particularly interesting, as it is localized in the westernmost tropical Pacific. Thus, it connects eastern and western Pacific while not leading to marked long-distance connections in the central Pacific close to the dateline. One appealing explanation of this finding could be that the corresponding link is mediated via the Walker circulation \citep{Ashok2009} and, thus, via airflow in higher altitudes rather than near-surface atmospheric circulation. However, it has to be noted that our analysis is based on cross-correlations only. The values can be severely affected by distinct temporal persistence properties of SAT in the eastern and western tropical Pacific, as pointed out by recent studies making use of modern causal inference methods \citep{Balasis2013,Runge2014}. Accounting for this effect in terms of replacing the correlation values by associated significance levels in the network generation step \citep{Palus2011} could provide a useful yet computationally demanding avenue for future research on this topic. From an impact perspective, the teleconnection suggested by our results is compatible with the documented increased likelihood of droughts in Indonesia during El Ni\~no events \citep{diaz2001enso}. 

The apparent teleconnection with Western Africa spans a rather large spatial distance (about one third of the globe). In this context, \citet{joly2009influence} noted that ''a significant part of the West African monsoon (WAM) interannual variability can be explained by the remote influence of El Niño–Southern Oscillation (ENSO).`` This previously reported teleconnection could be responsible for the elevated average link distance over Western Africa especially during EP El Ni\~nos. 

In general, climate variability in tropical regions is typically more likely to exhibit strong correlations than between tropics and extratropics, which is mainly due to the cellular structure of meridional atmospheric circulation that is effectively decoupling tropics and extratropics. In this regard, the omnipresent slightly elevated average link distance values in the polar regions are most likely data artifacts not corrected by our remapping procedure rather than indications of actual teleconnections.

\subsubsection{Regionalized network characteristics}

\begin{figure*}
	\centering
	\includegraphics[width=0.8\textwidth]{enso-colorbar.pdf} \\
	\subfloat{\includegraphics[width = \timeserieslen]{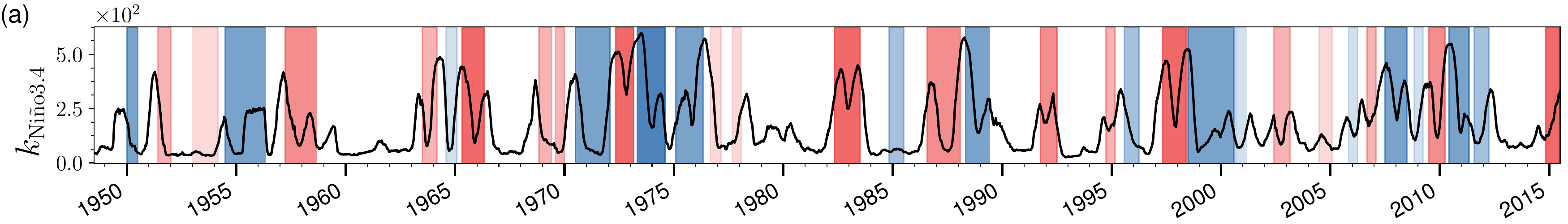}\label{fig:ts-degree-3.4}} \\
	\subfloat{\includegraphics[width = \timeserieslen]{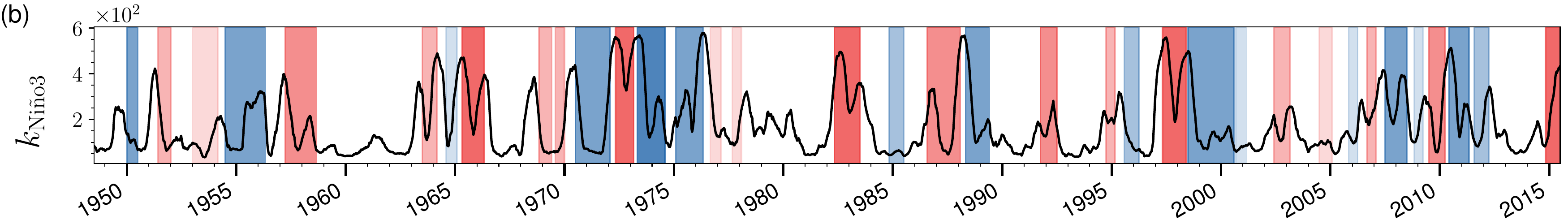}\label{fig:ts-degree-3}} \\
	\subfloat{\includegraphics[width = \timeserieslen]{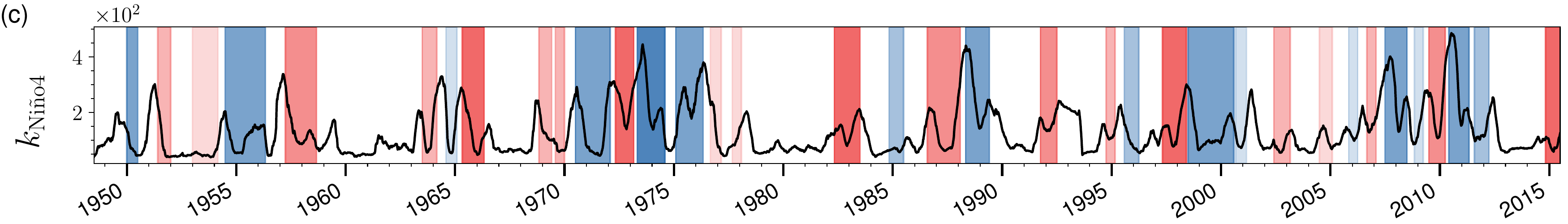}\label{fig:ts-degree-4}} \\
	\subfloat{\includegraphics[width = \timeserieslen]{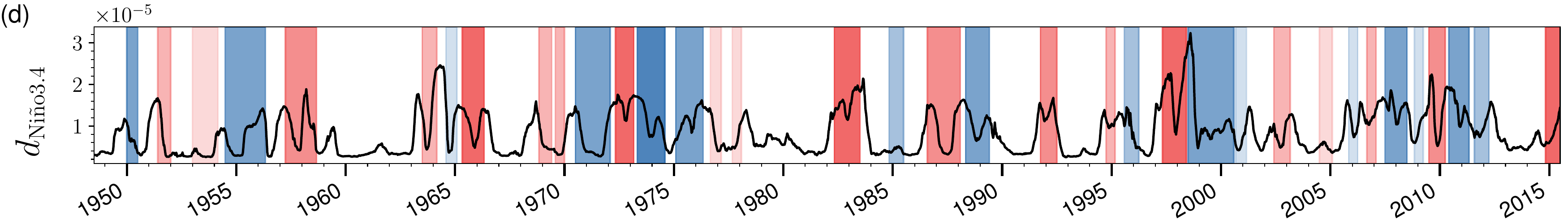}\label{fig:ts-link-length-3.4}} \\
	\subfloat{\includegraphics[width = \timeserieslen]{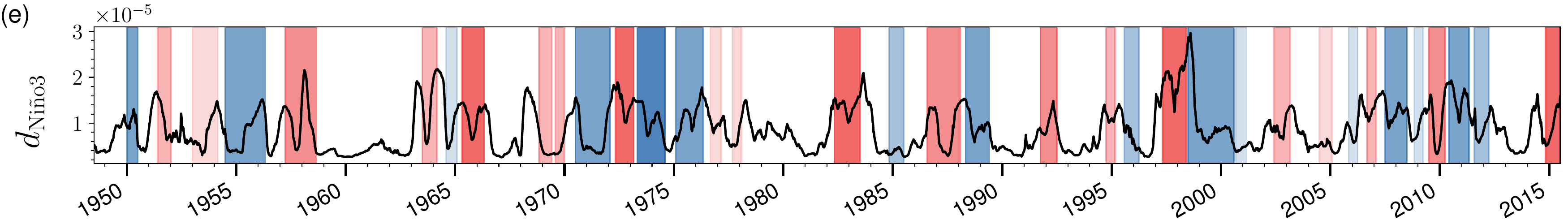}\label{fig:ts-link-length-3}} \\
	\subfloat{\includegraphics[width = \timeserieslen]{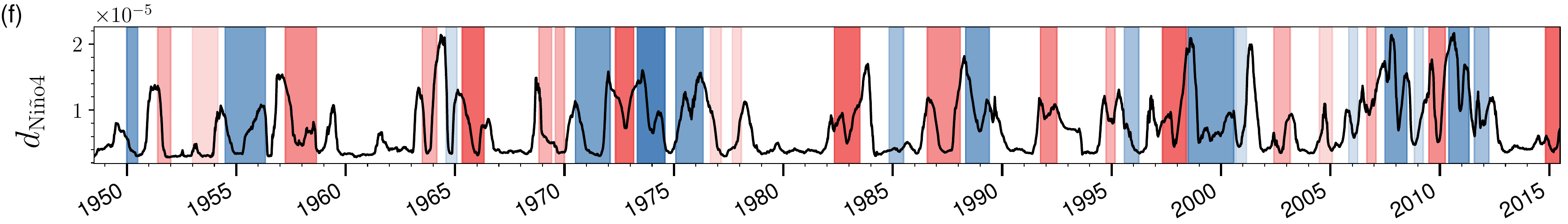}\label{fig:ts-link-length-4}} 
	\caption{Time series of different regionalized climate network properties introduced in Sect.~\ref{sec:localization-ENSO}. Background colours and time axis as in Fig.~\ref{fig:ts-global}.	}
	\label{fig:time-series-localized}
\end{figure*}

Global and local climate network properties as discussed above still provide only incomplete information on the effects of climate variability in different parts of the ENSO region on global SAT. To obtain further insights into this aspect, we now turn to analyzing the regionalized field measures introduced in Sect.~\ref{sec:localization-of-field-measures} and study the specific connectivity associated with the Nino3.4, Nino3 and Nino4 regions in terms of degree and average link distance.

The corresponding results are summarized in Fig.~\ref{fig:time-series-localized}. We observe that the relative magnitude of variations of regionalized degree and average link distance is even stronger than that of the global network properties transitivity, modularity and global average link distance discussed above. All measures exhibit episodes of very small values as opposed to such with much larger values, the latter often coinciding with El Ni\~no and La Ni\~na phases. Since the corresponding regions have been previously chosen for defining ENSO-specific indices, this result has been expected. Most importantly, degree and average link distance based characteristics exhibit strong positive correlations. Notably, for climatic events with predominantly local structure, we would expect a strong increase of $k_i$ but only weaker increase of $d_i$ in the region under study. Hence, our corresponding observations underline that ENSO-related climate impacts are not confined to the vicinity of the ENSO region, but are controlled by large-scale teleconnections.

Since the different ENSO regions show partial overlap (cf.~Fig.~\ref{fig:map-regions-of-interest}), the results obtained for the individual regions exhibit a high degree of similarity. However, regarding specific El Ni\~no or La Ni\~na episodes, comparing the corresponding signatures for the Nino3 and Nino4 regions still allows attributing these events to more Eastern Pacific or Central Pacific types. For example, the strong 1997/98 El Ni\~no is reflected by very high values of the regionalized degree for the Nino3 and Nino3.4 regions, but relatively weak signatures in the more western Nino4 region, which is consistent with its classification as an EP type event.

Examining the time evolution of all six regionalized network measures in some detail, it is notable that between 1978 and 1982, there has been considerable variability in all measures pointing towards an episodic presence of teleconnections even though none of the time windows was classified as an El Ni\~no or La Ni\~na episode according to the ONI. Moreover, we find that before the year 2000, clear peaks can always be observed in all properties as alternating with periods of low values. In turn, during the last about 15 years, we rather find strong variability without any low background level, with peaks occurring almost annually, with the exception of 2013 and 2014. This change in the overall temporal variability pattern of our regionalized network measures might point to some fundamental changes in the spatio-temporal organization of global SAT, either due to some not yet identified mode of natural variability or as a result of external interference. An attribution of this observation is, however, beyond the scope of the present work.

\subsection{Volcanic eruptions}
\label{sec:results-volcanoes}

Besides ENSO variability, strong volcanic eruptions have been identified as causes of marked disruptions in climate network properties in earlier studies \citep{radebach2013disentangling}. In this context, the application of the complementary viewpoints as used in this work for further characterizing the impacts of such eruptions promises interesting additional insights.

Regarding the global network properties, let us turn back to Fig.~\ref{fig:ts-global}. As already emphasized in our discussion on the corresponding imprints of different ENSO phases, anomalies in transitivity and modularity need to be interpreted differently in terms of global versus more regional changes in climate network connectivity. While EP El Ni\~nos and La Ni\~nas (but not their CP counterparts) consistently show peaks in transitivity coinciding with breakdowns in modularity, a similar signature has been found in the aftermath of the Mount Pinatubo eruption, suggesting that this event has affected the climate system globally. However, when comparing these topological network characteristics with the spatial network property of global average link distance $\left<\left<d\right>\right>$, we find a marked difference. Specifically, for ENSO-related climate disruptions, both $\left<\left<d\right>\right>$ and $\mathcal{T}$ show the same direction of changes (i.e., peaks) with only few exceptions. In turn, we observe an opposite behaviour of both measures in 1993, which corresponds to the time windows where the cooling effects following the Mount Pinatubo eruption should have taken their maximum \citep{McCormick1995}. Hence, unlike for ENSO-related disruptions, the peak in transitivity and simultaneous drop in $\left<\left<d\right>\right>$ indicate that the effects of the volcanic eruptions have rather been regionally confined. The latter is consistent with the hypothesis of elevated correlations in the region that has been most directly affected by the associated cooling trend following the eruption. Based on this observation, we suggest that using the global average link distance in conjunction with network transitivity and modularity enables us to discern disruptive events with global effects (strong ENSO phases) from those exhibiting more regional impacts (volcanic eruptions). 

\begin{figure*}
	\centering
	\subfloat{\includegraphics[width = \textwidth]{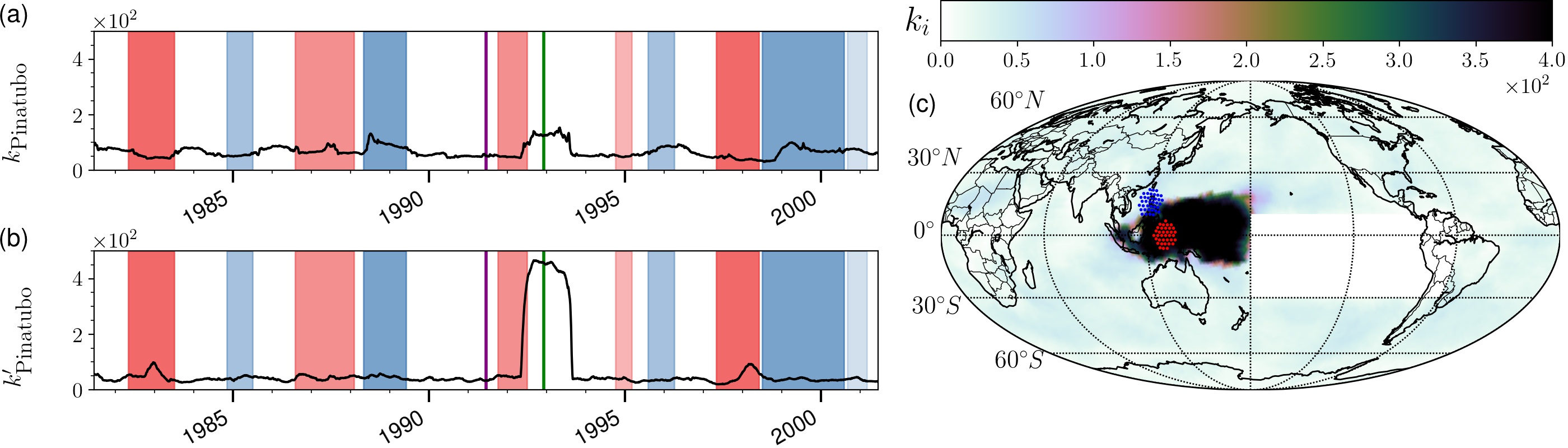}\label{fig:pinatubo-ts-degree}} \subfloat{\label{fig:pinatubo-ts-link-length}} \subfloat{\label{fig:pinatubo-map}} \\
	\subfloat{\includegraphics[width = \textwidth]{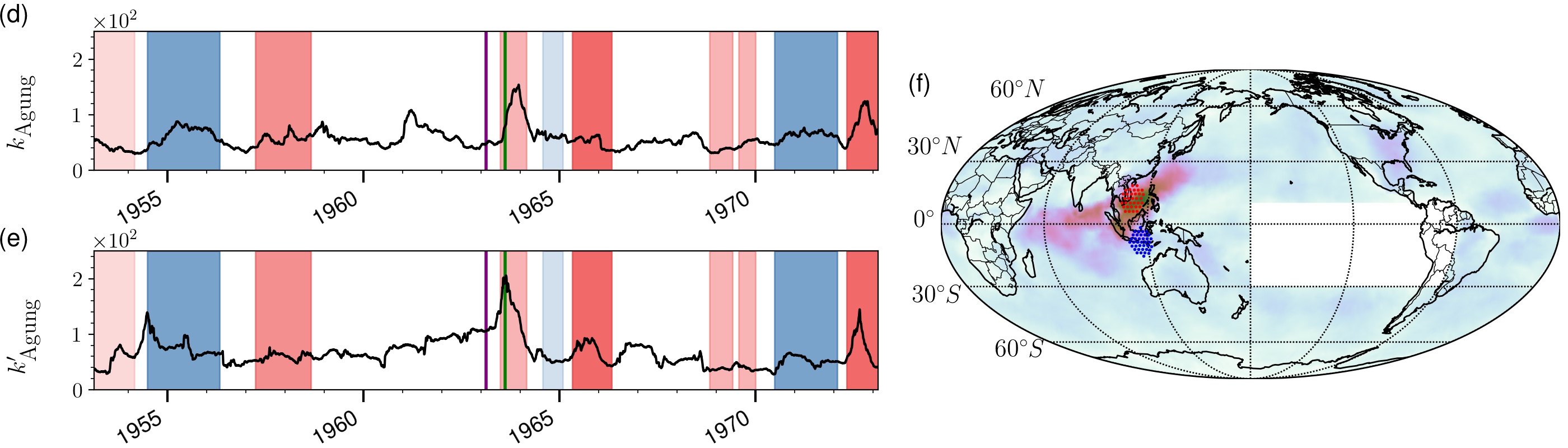}\label{fig:agung-ts-degree}} \subfloat{\label{fig:agung-ts-link-length}} \subfloat{\label{fig:agung-map}} \\
	\subfloat{\includegraphics[width = \textwidth]{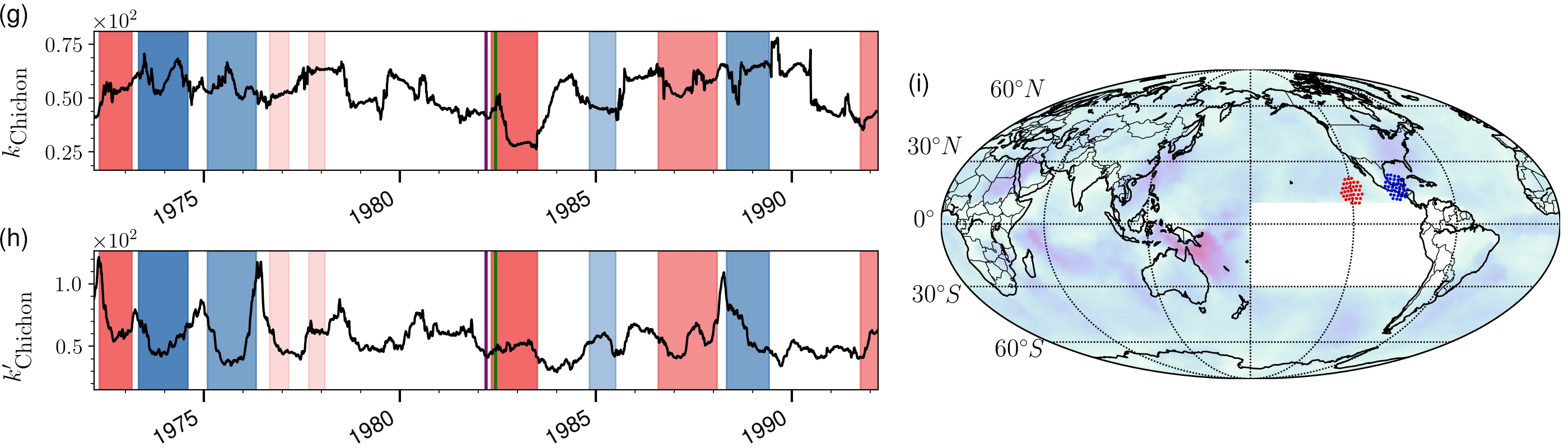}\label{fig:el-chichon-ts-degree}} \subfloat{\label{fig:el=chichon-ts-link-length}} \subfloat{\label{fig:el-chichon-map}} 
	\caption{Time series of regional mean degree and associated degree field (excluding the ENSO-big region indicated by white color) for the three main volcanic eruptions during the study period: (a-c) Mount Pinatubo, (d-f) Mount Agung, (g-i) El Chichon. In the degree maps shown in panels (c), (f) and (i), blue dots mark grid points within a radius of 5$\degree$ around each volcano, which have been used to define the regionalized degrees shown in panels (a), (d) and (g), respectively. Red dots indicate spatially shifted regions of the same size where the largest changes to the degree field have been observed. These regions serve as the basis for computing the regionalized degrees shown in panels (b), (e) and (h), respectively. Purple vertical lines indicate the timing of the respective eruptions, whereas green vertical lines indicate the midpoints of the time windows exhibiting the strongest signature in the regionalized network properties. The time series have been restricted to $\pm$10 years around the date of the respective eruption. Background colours indicate the corresponding ENSO strength as in Fig.~\ref{fig:ts-global}.}
	\label{fig:volcanoes}
\end{figure*}

In general, one notable difference in comparison with ENSO-related impacts is that large-scale effects of volcanic eruptions on global SAT teleconnectivity can be observed only after a sufficiently large amount of aerosols have entered the stratosphere \citep{Robock2000}. Accompanying the resulting time shift between trigger event and response, we may also need to consider a spatial shift of the most affected region as compared to the location of the volcano. In the following, we apply our regionalization procedure described in Sect.~\ref{sec:localization-volcanoes} to studying the impacts of the Mount Pinatubo, Mount Agung and El Chichon eruptions. In order to avoid interference with the effects of ENSO events, the ENSO-big region depicted in Fig.~\ref{fig:map-regions-of-interest} are excluded from the corresponding computations. The results obtained from this analysis are summarized in Fig.~\ref{fig:volcanoes}.

The largest of the three considered eruptions (Mount Pinatubo) had a global cooling effect and has left clearly visible signatures in all considered global network measures as discussed above. Some months after the eruption, a large region of elevated network connectivity has been established, which covers essentially all of the western tropical Pacific (Fig.~\ref{fig:pinatubo-map}). The temporal evolution of the average degree in the region around Mount Pinatubo displays an abrupt rise about half a year after the eruption, then a constantly high value for about one year (the common residence time of volcanic aerosols in the stratosphere) before dropping again back to its previous level (Fig.~\ref{fig:pinatubo-ts-degree}). The region with the highest degrees is shifted northward with respect to the location of the volcano (Fig.~\ref{fig:pinatubo-map}). When computing the average degree for this region, we observe an even stronger rise of the regionalized degree than for the region surrounding the volcano (Fig.~\ref{fig:pinatubo-ts-link-length}). 

The Mount Agung eruption exhibits similar, but weaker, patterns in the respective region (Fig.~\ref{fig:agung-map}). However, the region with the highest degree is shifted south-westward. The average degree in the region surrounding Mount Agung only shows weak changes after the eruption (Fig.~\ref{fig:agung-map}), as opposed to a somewhat sharper increase in the shifted region, with the peak effect occurring significantly faster after the beginning of the eruption than in case of the Mount Pinatubo eruption (Fig.~\ref{fig:agung-ts-link-length}). However, it should be noted that we can already observe the beginning of some upward trend in regionalized degree before the actual eruption, pointing to a possible interference with normal natural variability. 

Unlike the two other volcanic eruptions, the degree field in the period succeeding the El Chichon eruption showed hardly any marked changes (Fig.~\ref{fig:el-chichon-map}). Consequently, we also do not observe any marked signature in the temporal variability profile of the regionalized degree in the surrounding of the volcano (Fig.~\ref{fig:el-chichon-ts-degree}). Instead of a peak shortly after the eruption, we actually find a clear drop of the average degree. However, given that El Chichon is located relatively close to the extended ENSO region, we cannot rule out that this could be an effect of the strong El Ni\~no event occurring shortly after the eruption and eventually even being partially triggered by the latter \citep{Khodri2017}. In general, previous studies indicate that the El Chichon eruption caused a much weaker summer cooling than the Mount Agung eruption \citep{Man2014}, which could also explain its absent signature in our analysis.

\section{Conclusions}

We have used functional climate networks constructed from spatial correlations of daily global surface air temperature (SAT) anomalies to analyze the global impact and teleconnections of past El Niño and La Niña events as well as volcanic eruptions. By making use of the global network property of modularity, we have found that at least the East Pacific flavours of such events lead to a \emph{global reconfiguration} of SAT variations. Considering the global average link distance as a complementary spatial network characteristic, we have identified distinct qualitative differences between the imprints of these ENSO periods and the Mount Pinatubo eruption in global SAT patterns.

Using composites of the degree and average link distance fields, we have identified hallmarks of distinct ENSO teleconnections in the climate network structure, especially such linking the eastern tropical Pacific with Indonesia and West Africa during East Pacific El Ni\~nos, both of which have already been reported in previous studies \citep{diaz2001enso,joly2009influence}. By making use of a regionalization procedure applied to these two fields of local network properties, we have introduced a simple yet effective tool to unveil the differential roles of different regions in the tropical Pacific in establishing teleconnections during different El Ni\~no and La Ni\~na events.

Finally, we have analyzed the global and local connectivity properties of SAT-based climate networks in the aftermath of the strongest recent volcanic eruptions of Mount Pinatubo, Mount Agung and El Chichon. In particular, while the Mount Pinatubo eruption has been confirmed to exhibit marked impacts on global SAT, its dominating effect was rather regional (i.e., it did not trigger long-range teleconnections detectable by our approach).

While most of the results presented in this work rely on a qualitative analysis of temporal changes in the climate network properties, additional statistical quantification of their relationship with existing indicators of ENSO variability and volcanic eruptions' impacts might further strengthen our findings. Regarding ENSO, many previous studies have attempted to utilize the spatial patterns of SST anomalies to define corresponding index variables. However, the corresponding classifications of El Ni\~no and La Ni\~na phases reached only partial consensus, which was in fact the motivation of the work of \citet{Wiedermann2016a} presenting climate network transitivity as a useful and consistent index. Going one step further, one might easily quantify, for example, the correlation between transitivity and other (global or regionalized local) network characteristics. However, in our opinion, the particular value of the present work lies in identifying specific properties that are not exhibited by the former (as well as other not network-related indices). In this context, there is no established benchmark that could be used for further testing the significance of our results. 

In turn, regarding the effects of volcanic eruptions, the respective regionalized degrees for the spatially shifted ``major impact regions'' of both, Mount Pinatubo and Mount Agung, exhibited their overall maximum values among all time windows studied in this work in the aftermath of the associated eruptions. This indicates a very high significance of our corresponding results. Note that, however, we did not succeed in finding any comparatively strong impact signature in the climate network properties after the eruption of El Chichon, as well as after other strong volcanic eruptions of the past about 70 years (not shown). We relate the latter finding to the generally lower magnitude of the respective perturbations (in terms of a smaller amount of climate-active volcanic aerosols injected into the stratosphere). Moreover, some of the other major eruptions (e.g., the Mount St.~Helens eruption in 1980) appeared in the extratropics rather than the tropics. Together with the different seasonality of these events, this could imply different effects on regional and global temperature patterns, similar as shown recently for global monsoon precipitation \citep{Liu2016}.

In summary, our study confirms that ENSO does not only have a strong local effect on SAT in terms of coherent SAT trends in the tropical Pacific associated with a spatially confined increase of network connectivity \citep{radebach2013disentangling}, but also dynamically reconfigures climate variability globally by triggering teleconnections especially with other tropical regions. In this regard, one possible mechanism could involve the modulation of monsoons by strong El Ni\~no and/or La Ni\~na periods, which could be further modulated by volcanic eruptions \citep{Maraun2005}. Confirming this hypothesis in the context of climate network studies would, however, require much more elaborated approaches than those used in the present work, and is therefore outlined as a subject of future research.

\appendix

\section{Comparison of different modularity estimation algorithms}
\label{app:modularity-comparison}

In Fig.~\ref{fig:modularity-comparison}, we show the results of five algorithms to estimate the community structure of our functional climate networks in terms of the resulting modularity values: fast greedy \citep{clauset2004finding}, infomap \citep{rosvall2008maps}, label propagation \citep{raghavan2007near}, leading eigenvector \citep{newman2006finding} and WalkTrap \citep{pons2006computing}. Further details motivating the choice of WalkTrap as a reference algorithm in the body of this paper are provided in the figure caption.

\begin{figure*}
	\centering
	\includegraphics[width=\textwidth]{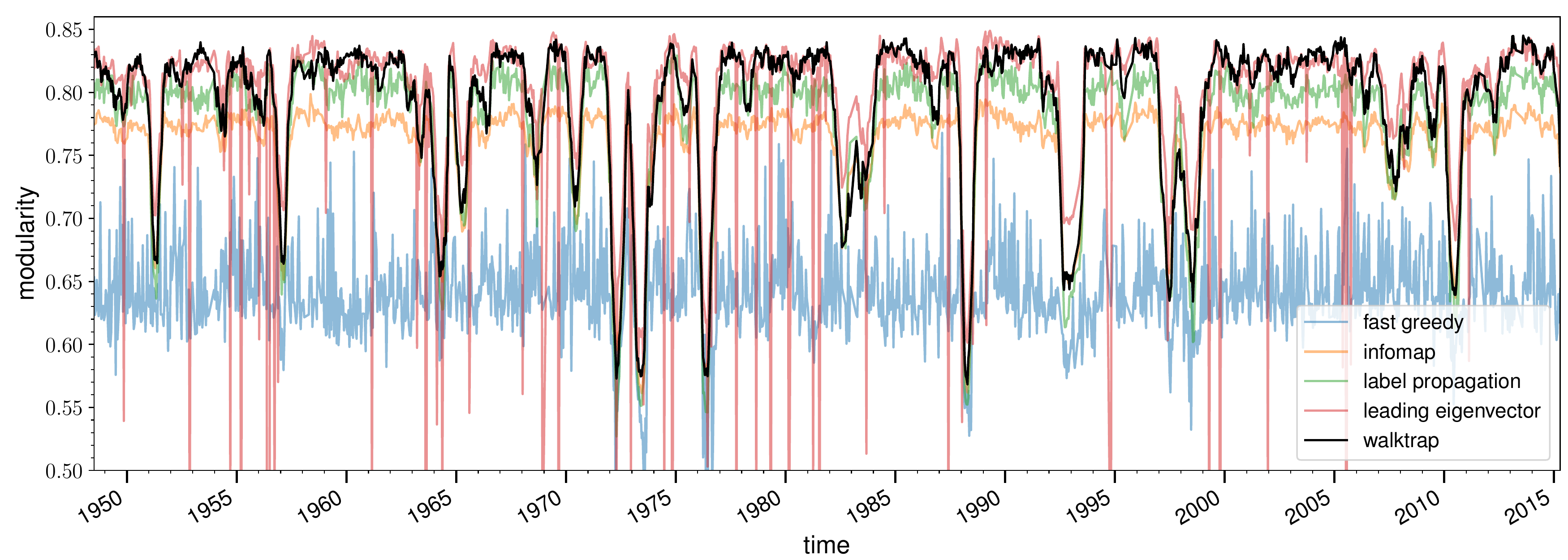}
	\caption{
		Comparison of estimated modularity values for the functional climate networks obtained for running windows as described in the main text.
		We use five different algorithms for detecting the underlying community structure.
		Since modularity estimation resorts to a numerical maximization problem, higher values indicate better results.
		Visual comparison reveals that the leading eigenvector and WalkTrap algorithms outperform the others regarding this criterion.
		Since the leading eigenvector algorithm suffers from intermittent modularity breakdowns, possibly indicating numerical instabilities, we use the WalkTrap method in this paper.}
	\label{fig:modularity-comparison}
\end{figure*}

\vspace{0.3cm}
\noindent
\emph{Data availability.} All data used in this work are publicly available from NCEP/NCAR. In addition to the SAT data, we have made use of the monthly Ocean Ni\~no Index (ONI) values as provided by \citet{NCEP-ONI}.

\vspace{0.3cm}
\noindent
\emph{Code availability.} All code used in this work has been written in Python and published under GPLv3 license as a GitHub repository \citep{kittel-code}. Detailed information for the reproduction of the results of this paper can be found there. While the published code was originally designed to produce these specific result, we kept it rather general with further extensions in mind. Thereby, it can be used as a starting point for future evolving network research, as it provides some basic structures that are needed for evolving network analysis, for example an interface for \texttt{hdf5} (a high-performant file type for data storage) and automatic parallelization using \texttt{mpi}.

\authorcontribution{TK, CC, NL, FR and RVD designed the analysis. TK, CC and NL conducted the analysis. TK, CC and RVD prepared the manuscript. TP, FR, JK and RVD supervised the analysis and revised the manuscript and the interpretation of the obtained results.}

\competinginterests{The authors declare no conflict of interest.}

\begin{acknowledgements}

This work has been financially supported by the International Research Training Group IRTG 1740/TRP 2014/50151-0, jointly funded by the German Research Foundation (DFG, Deutsche Forschungsgemeinschaft) and the S\~{a}o Paulo Research Foundation (FAPESP, Funda\c{c}\~{a}o de Amparo \`a Pesquisa do Estado de S\~{a}o Paulo) and by the German Federal Ministry for Education and Research (BMBF) via the Young Investigators Group "CoSy-CC$^2$: Complex Systems Approaches to Understanding Causes and Consequences of Past, Present and Future Climate Change" (grant no. 01LN1306A) and the JPI Climate/Belmont Forum project "GOTHAM: Globally Observed Teleconnections and Their Representation in Hierarchies of Atmospheric Models".
This work was conducted in the framework of PIK’s flagship projects on coevolutionary pathways (\textsc{copan}), time series analysis (\textsc{tsa}) and networks of networks (\textsc{neonet}).
The authors thank
Nikoo Ekhtiari,
Jobst Heitzig,
Finn M\"uller-Hansen,
Stefan Ruschel,
and Marc Wiedermann
for helpful discussions.
Furthermore, the European Regional Development Fund (ERDF), the German Federal Ministry of Education and Research and the Land Brandenburg are acknowledged for supporting this project by providing resources on the high performance computer system at the Potsdam Institute for Climate Impact Research.
Finally, the authors wish to express their thanks to the developers of Python \citep{python} the used packages \texttt{numpy} (Numerical Python, \citet{numpy}), \texttt{scipy} (Scientific Python, \citet{scipy}), \texttt{matplotlib} \citep{matplotlib,matplotlib-publication} and \texttt{igraph} for Python \citep{igraph,igraph-citation}.

\end{acknowledgements}

\bibliographystyle{copernicus}
\bibliography{network-communities}

\end{nolinenumbers}
\end{document}